\documentclass[printer]{aa}
\usepackage{natbib,graphicx}
\usepackage{txfonts}

\begin{document}
\title{The 9.7 and 18 $\mu$m silicate absorption profiles towards diffuse and molecular cloud lines-of-sight}
\author{J.M. van Breemen\inst{1} \and M. Min\inst{1,2} \and J.E. Chiar\inst{3} \and L.B.F.M. Waters\inst{1,4,5} \and F. Kemper\inst{6,7} \and A.C.A. Boogert\inst{8} \and J. Cami\inst{9,3} \and L. Decin\inst{4,1} \and C. Knez\inst{10} \and G.C. Sloan\inst{11} \and A.G.G.M. Tielens\inst{12}}

\institute{Astronomical Institute ``Anton Pannekoek'', University of
  Amsterdam, Science Park 904, 1098 XH Amsterdam, The Netherlands \and
  Astronomical Institute Utrecht, University of Utrecht, PO Box 80000,
  3508 TA Utrecht, The Netherlands \and SETI Institute, 515 North
  Whisman Road, Mountain View, CA 94043, USA \and Department of
  Physics and Astronomy, Institute for Astronomy, K.U.~Leuven,
  Celestijnenlaan 200B, 3001 Leuven, Belgium \and SRON Netherlands
  Institute for Space Research, Sorbonnelaan 2, 3584 CA Utrecht, The
  Netherlands \and Jodrell Bank Centre for Astrophysics, Alan Turing
  Building, School of Physics and Astronomy, The University of
  Manchester, Oxford Road, Manchester, M13 9PL, UK \and Academia
  Sinica Institute of Astronomy and Astrophysics, P.O. Box 23-141,
  Taipei 10617, Taiwan \and IPAC, NASA Herschel Science Center, Mail
  Code 100-22, California Institute of Technology, Pasadena, CA 91125,
  USA \and Physics and Astronomy Department, University of Western
  Ontario, London ON N6A 3K7, Canada \and Department of Astronomy,
  University of Maryland, College Park, MD 20742, USA \and Cornell
  University, Astronomy Department, 108 Space Sciences Bldg., Ithaca,
  NY 14853-6801, USA \and Leiden Observatory, PO Box 9513, 2300 RA
  Leiden, The Netherlands}

   \date{Received / Accepted }

   \abstract {Studying the composition of dust in the interstellar
     medium (ISM) is crucial in understanding the cycle of dust in our
     galaxy.}  {The mid-infrared spectral signature of {amorphous
       silicates, the most abundant dust species in the ISM,} is
     studied in different lines-of-sight through the {Galactic} plane,
     thus probing different conditions in the ISM.}  {We have analysed
     10 spectra from the Spitzer archive, of which 6 lines-of-sight
     probe diffuse interstellar medium material and 4 probe molecular
     cloud material. The 9.7 $\mu$m silicate absorption features in 7
     of these spectra were studied in terms of their shape and
     strength. In addition, the shape of the 18 $\mu$m silicate
     absorption features in 4 of the diffuse sightline spectra were
     analysed.}  {The 9.7 $\mu$m silicate absorption {bands} in
     the diffuse sightlines show a strikingly similar band shape. This
     is also the case for all but one of the 18 $\mu$m silicate
     absorption bands observed in diffuse lines-of-sight. The 9.7
     $\mu$m {bands} in the 4 molecular sightlines show small
     variations in shape. These modest variations in the band shape
     are inconsistent with the interpretation of the large variations
     in $\tau_{9.7}$/$E$(J$-$K) between diffuse and molecular
     sightlines in terms of silicate grain growth. Instead, we suggest
     that the large changes in $\tau_{9.7}$/$E$(J$-$K) must be due to
     changes in $E$(J$-$K).}{}

   
   \authorrunning{J.M.~van Breemen et al.}
   \titlerunning{Silicate absorption profiles in molecular and diffuse interstellar sightlines}

   \maketitle

\section{Introduction}

The composition of dust in the interstellar medium (ISM) is a result
of {a variety of} different processes. First of all, dust is formed in
the circumstellar environments of evolved stars, for example
asymptotic giant branch (AGB) stars. Depending on their stage of
evolution these stars produce either oxygen-rich (e.g. silicates and
oxides) or carbon-rich dust (e.g. amorphous carbon, {graphite
  and silicon carbide}). The {newly-formed} {stardust} enters
the ISM through stellar winds or supernova explosions. This material
is rapidly mixed with other gas and dust in the ISM, where it is
processed by a number of mechanisms {such as} shock waves driven by
supernova explosions, high energy radiation and high velocity
collisions amongst grains. Finally, during the formation of a new star
and planetary system the dust is further processed. Therefore, the
composition of dust in the ISM is a reflection of the constant
formation, mixing, processing and destruction of different dust
species. Studying the composition of interstellar dust and its spatial
variations throughout the galaxy is crucial to get a better
understanding of the material from which, for instance, our solar
system was made.

A good way to study the composition of interstellar dust is by means
of infrared spectroscopy. The light {from the bright background star
  is attenuated} by the interstellar dust in front of it. This
extinction is wavelength dependent and is caused by scattering as well
as absorption by the dust grains. If the intrinsic, unreddened
spectrum of the background star is known, the dust extinction as a
function of wavelength can be derived.

\begin{table*}
\hspace*{-0.5cm}
\begin{tabular}{llllllll}
\hline \\ 
target name & {other} & l & b & Program ID & spect.& diffuse (D) or & 9.7 or 18 $\mu$m\\
 & {designations} & (degrees) & (degrees) & & type & molecular (M) & silicate feature\\
\hline\\
StRS 136 & {2MASS J17475608} & 0.040937 & $-$0.566892 & 3616 & B8--A9I$^{2}$ & D & 9.7\\
& {$-$2911439} & & & & & & \\
StRS 164 & {2MASS J18161876} & 14.213568 &  $-$0.002043 & 3616 & B8--A9I$^{2}$ & D & 9.7\\
& {$-$1635468} & & & & & & \\
{SSTc2d\_J182835.8+002616} & {\ldots} & 30.712112 & 5.280682 & 139 $\&$ 172-179 & \ldots & M (Serpens) & 9.7\\
{$[$SVS76$]$ Ser 9}$^{1}$ & {2MASS J18294508}  & 31.626616 & 5.423514 & 139 $\&$ 172-179 & \ldots & M (Serpens) & 9.7\\
& {+0118469} & & & & & & \\
StRS 354 & {IRAS 20273+3740} & 76.972032 & $-$0.635686 & 20294 & O7--B3$^{2}$ & D & 9.7 \& 18\\
{Elias 3-13}$^{3}$ & {IRAS 04303+2609} & 172.694191 & $-$14.498000 & 139 $\&$ 172-179 & K2III$^{3}$ & M (Taurus) & 9.7\\
SSTc2d\_J163346.2$-$242753 & {\ldots} & 354.158006 & 15.606477 & 139 $\&$ 172--179 & \ldots & M ($\rho$ Ophiuchi) & 9.7 \\
{G323.2103$-$00.3473}$^{4}$ & {2MASS J15285631} & 323.211041 & $-$0.347425 & 3616 & \ldots & D & 18\\
& {$-$5653045} & & & & & & \\
{G343.6142$-$00.1596}$^{4}$ & {2MASS J16585873} & 343.613830 & 0.160012 & 3616 & \ldots & D & 18\\
& {$-$4220543} & & & & & & \\
{G345.3650$-$00.4015}$^{4}$ & {2MASS J17070616} & 345.364763 & $-$0.401650 & 3616 & \ldots & D & 18\\
& {$-$4118140} & & & & & & \\
\hline\\
\end{tabular} 
\caption{Overview of the selected spectra from the Spitzer archive. Listed are the target name and the {Galactic} longitude {and latitude}. The Spitzer ID of the program that contains the observation and the spectral type of the source {is} listed and it is indicated if the observation was done in a diffuse (D) or molecular (M) sightline. In the case of molecular sightlines, the corresponding molecular cloud is listed. Finally, we list whether the 9.7 or 18 $\mu$m silicate feature was extracted from these spectra. $^{1}$\citet{1976AJ.....81..314S}, $^{2}$\citet{Rawlings00}, $^{3}$\citet{Elias78}, $^{4}$\citet{2003yCat.5114....0E}.}
\label{tab:samplesel}
\end{table*}

The most abundant dust species in the ISM are amorphous silicates,
which cause two prominent absorption features at about 9.7 and 18
$\mu$m
\citep[e.g.~][]{1971Natur.233...72S,1974ApJ...193L..81R,1975A&A....45...77G,1980ApJ...242..965M,1989ESASP.290...79R,1998MNRAS.298..131B,Kemper04,Chiar06,Min07}.
{In this article we will discuss both the 9.7 $\mu$m and 18 $\mu$m
  bands.}  Interstellar silicate dust is mostly a mixture of amorphous
silicates with an olivine (O/Si=4) or pyroxene (O/Si=3) composition
\citep[e.g.][]{1974ApJ...192L..15D, 1975A&A....45...77G, Kemper04,
  Min07, Chiar07}. Although olivines {and} pyroxenes are by definition
crystalline silicates, in this paper we will also use these names for
amorphous silicates {with the same stoichiometric} composition. From
previous studies we know that the composition of interstellar
silicates can vary, depending on the location in the ISM. For example
\citet{Demyk99,Demyk00,Demyk01} showed that the dust around protostars
has a higher O/Si ratio than the {newly-formed} dust observed around
AGB stars.

\citet{Chiar07} compared the strength of the 9.7 $\mu$m silicate
absorption feature with the near-infrared colour excess $E$(J$-$K) in
both diffuse and molecular sightlines\footnote[1]{Diffuse and
  molecular sightlines are defined as lines-of-sight which pass
  mostly, but not exclusively, through diffuse or molecular cloud
  material, respectively.}. {The 9.7 $\mu$m silicate feature,
  $\tau_{9.7}$, is due to the Si-O stretching mode in the silicates,
  and its strength depends on the silicate optical depth, while the
  overall extinction in the visible through {near-infrared} is
  probably due to a combination of carbonaceous material
  \citep[e.g.~amorphous carbon;][]{Draine84,CVJ_10_dustSED} and iron,
  the latter either in metallic form \citep{KDW_02_composition} or as
  a cationic part of the silicate lattice, the so-called \emph{dirty
    silicates} \citep{JM_76_dust}}.  For diffuse sightlines there is a
tight linear correlation between $\tau_{9.7}$ and the visual
extinction ($A_{\mathrm{V}}$) \citep{Roche84,Whittet03}. A similar
correlation exists between $\tau_{9.7}$ and the near-infrared
extinction represented by the J-K colour excess, $E$(J$-$K)
\citep{Chiar07}. However, for the molecular sightlines the diffuse ISM
correlation fails \citep{Whittet88_2,Chiar07}, indicating that
something happens to the dust grains when they enter a molecular
cloud. Molecular sightlines, in general, probe {greater interstellar
  densities} than diffuse sightlines and thus, the observed variations
are most likely density related.

An obvious explanation is grain growth via coagulation taking place in
molecular clouds. Indeed, this causes a decrease in the strength of
the 9.7 $\mu$m silicate feature (see Sect.~\ref{sect:models}), but
grain growth also has an effect on the {near-infrared} extinction
(i.e. $E$(J$-$K)). Moreover, {grain growth} alters the \emph{shape} of
the 9.7 $\mu$m silicate feature drastically
{\citep[e.g.~][]{BMD_01_processing}}.  {Therefore, an accurate
  measurement of the band shape of the 9.7 $\mu$m silicate band in
  diffuse and molecular lines-of-sight can set interesting limits on
  the importance of grain growth as an explanation for the large
  changes in extinction detected so far.}

In order to study the spectral signature of interstellar silicates in
both diffuse and molecular sightlines, we have analysed 10
low-resolution spectra {with high S/N ratio,} taken with the Spitzer
space telescope.  {The high quality of the spectra allows an accurate
  comparison of band shapes along different lines-of-sight at a
  variety of Galactic longitudes}.  Of these spectra, 7 were used to
analyse the shape and strength of the 9.7 $\mu$m silicate absorption
feature and 4 were used to analyse the 18 $\mu$m silicate absorption
profile. {Only a few earlier studies have analysed the interstellar 18
  $\mu$m band, and always in conjunction with the 9.7 $\mu$m band
  \citep[e.g.~][]{Demyk99,Chiar06}.}

This paper is organised as follows: Section 2 describes the sample
selection and the observations. In Sect.~3 the data analysis is
described and the results are presented. {These results are
  compared with model calculations in Sect.~4, and in Sect.~5 we
  discuss the implications of our results for grain processing in
  molecular clouds. Finally, the conclusions are summarised in
  Sect.~6.}

\section{Observations}

\subsection{Sample selection}
Our sample consists of a selection of low-resolution IRS
{\citep[InfraRed Spectrograph;][]{2004ApJS..154...18H}} Spitzer
spectra from the Spitzer programs with Program {IDs}: 139,
172--179, 3616, 20294, 3336, 20604. These programs have, amongst other
things, observed background sources with large amounts of foreground
dust and were thus relevant for this study. {The background
  nature of the dense cloud sources was based on their placement in
  colour-colour diagrams which separated out stars reddened by
  interstellar, rather than circumstellar dust. For the diffuse ISM
  lines-of-sight, the background nature is confirmed via their strong
  reddening and optical spectra that confirm their spectral and
  luminosity class \citep{Rawlings00}.}

A first selection was made based on the presence and strength of the
9.7 $\mu$m and 18 $\mu$m silicate absorption features and the signal
to noise ratio (SNR) of the spectra. From the resulting list 10
spectra were selected keeping in mind the following science goals: (1)
Investigating the optical depth at 9.7 $\mu$m, $\tau_{9.7}$, versus
the infrared colour excess, $E$(J$-$K), (2) Comparing the shape of the
9.7 $\mu$m silicate features through different sightlines, (3)
Comparing the shape of the 18 $\mu$m silicate features through
different sightlines.

\begin{figure}[!h]
\centering
\includegraphics[width=6cm,angle=270]{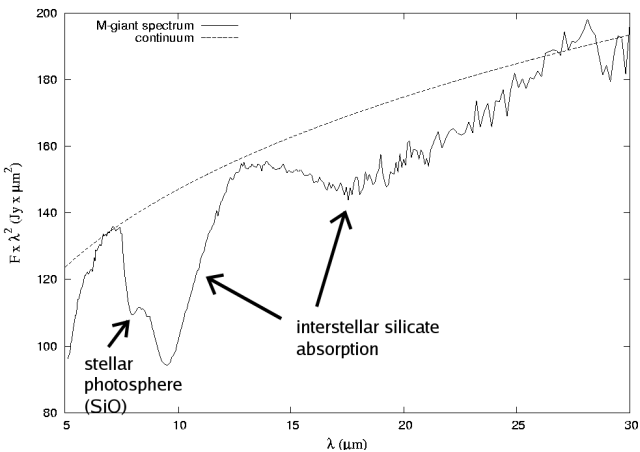}
\caption{Spectrum of an M-giant taken with the Spitzer Space Telescope
  (Chiar, unpublished Spitzer data). The extinction towards this star
  is $A_{\mathrm{V}}$ $\sim$ 10 mag. Plotted is the observed flux (in Jy) of the
  star multiplied by the wavelength (in $\mu$m) squared ($F_{\nu} \times
  \lambda^{2}$) versus the wavelength ($\lambda$). {This choice
    implies that a horizontal line is expected for an unreddened
    photosphere.} The two most prominent absorption bands at 9.7 and
  18 $\mu$m are due to interstellar silicate absorption. The smaller
  band at 8 $\mu$m is caused by gas phase SiO in the photosphere of
  the M-giant itself and clearly affects the 9.7 $\mu$m silicate
  feature, thus hampering the detailed analysis of the shape of the
  interstellar silicate band.}
\label{fig:Mspectrum}
\end{figure}

To reach the first two science goals the 9.7 $\mu$m silicate feature
needs to be extracted from the spectrum. However, many of the spectra
of background stars in the Spitzer archive show an SiO gas phase
fundamental stretch band from about 8 to 11 $\mu$m coming from the
intrinsic spectrum of the observed background {star} (see
Fig.~\ref{fig:Mspectrum}). This photospheric gas phase SiO band is
typical for late type K and M giants and because these are
intrinsically bright {stars}, they are good candidates for
background sources.

Since the shape of the SiO band depends on the spectral type of the
background source, which in most cases is not known,
{accurately correcting for this band is not easy (see also
  Sect.~\ref{sec:molsight}). The profile of the SiO fundamental mode
  is highly sensitive to the temperature and the structure of the
  stellar atmosphere, requiring modelling beyond that of a classical
  atmosphere \citep[e.g.~][]{1994A&A...289..469T}.} Therefore, 7
spectra were selected that show no {or} very little SiO {absorption}. Of
these 7 spectra 3 are diffuse sightlines and 4 are molecular
sightlines. All but one of these spectra were not suited to analyse
the 18 $\mu$m silicate absorption feature, either because the spectrum
did not span the full spectral range or the quality of the long
wavelength part of the spectrum was not high enough to reliably
extract the 18 $\mu$m feature. Therefore, we have selected 3 more
spectra from the Spitzer archive to analyse the 18 $\mu$m silicate
absorption feature, which in turn were not suited for the analysis of
the 9.7 $\mu$m feature, because they show strong photospheric gas
phase SiO bands (see Fig.~\ref{fig:Mspectrum}). Unfortunately, we did
not find any molecular sightline spectra in the Spitzer archive that
span the long wavelength part of the Spitzer spectral range and
therefore, we could only analyse the 18 $\mu$m silicate absorption
feature in 4 diffuse sightlines. An overview of the selected spectra
is given in Table~\ref{tab:samplesel}.

\subsection{Data reduction}

\begin{figure*}
\centering
\includegraphics[angle=270,width=5cm]{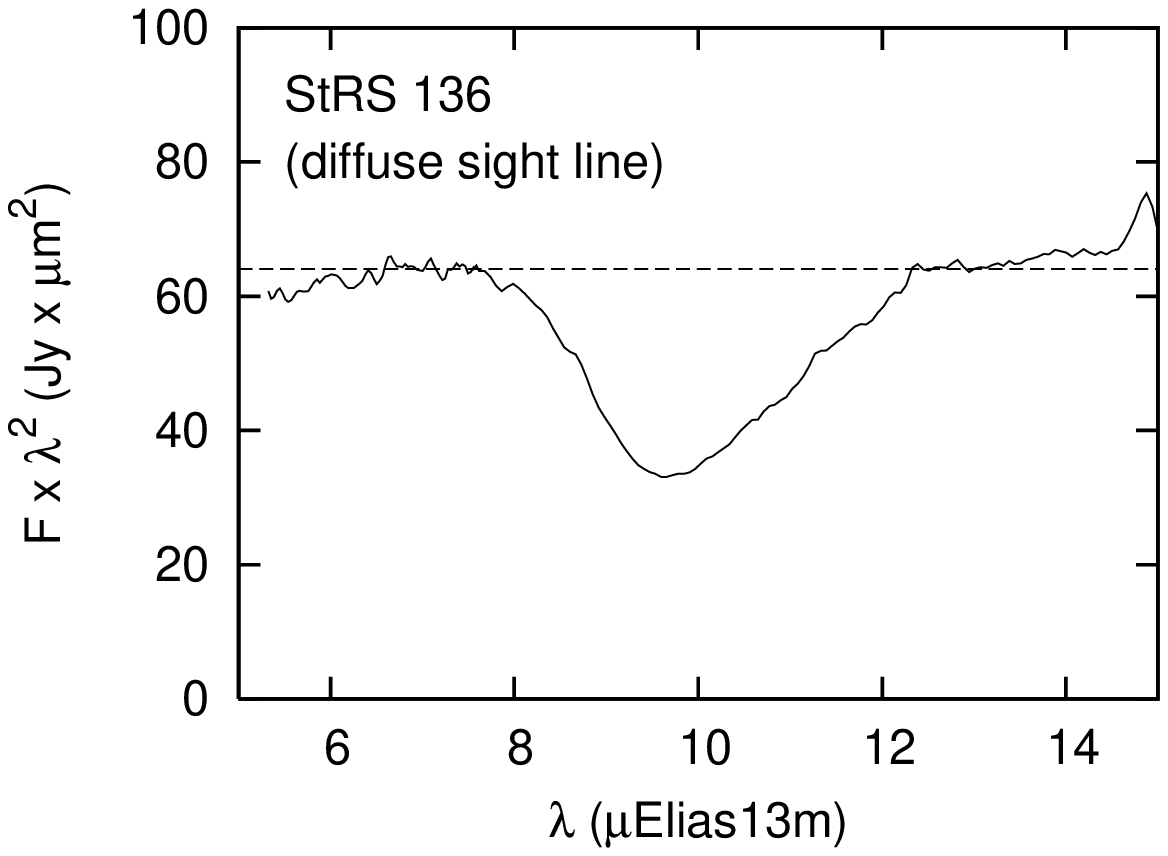}
\includegraphics[angle=270,width=5cm]{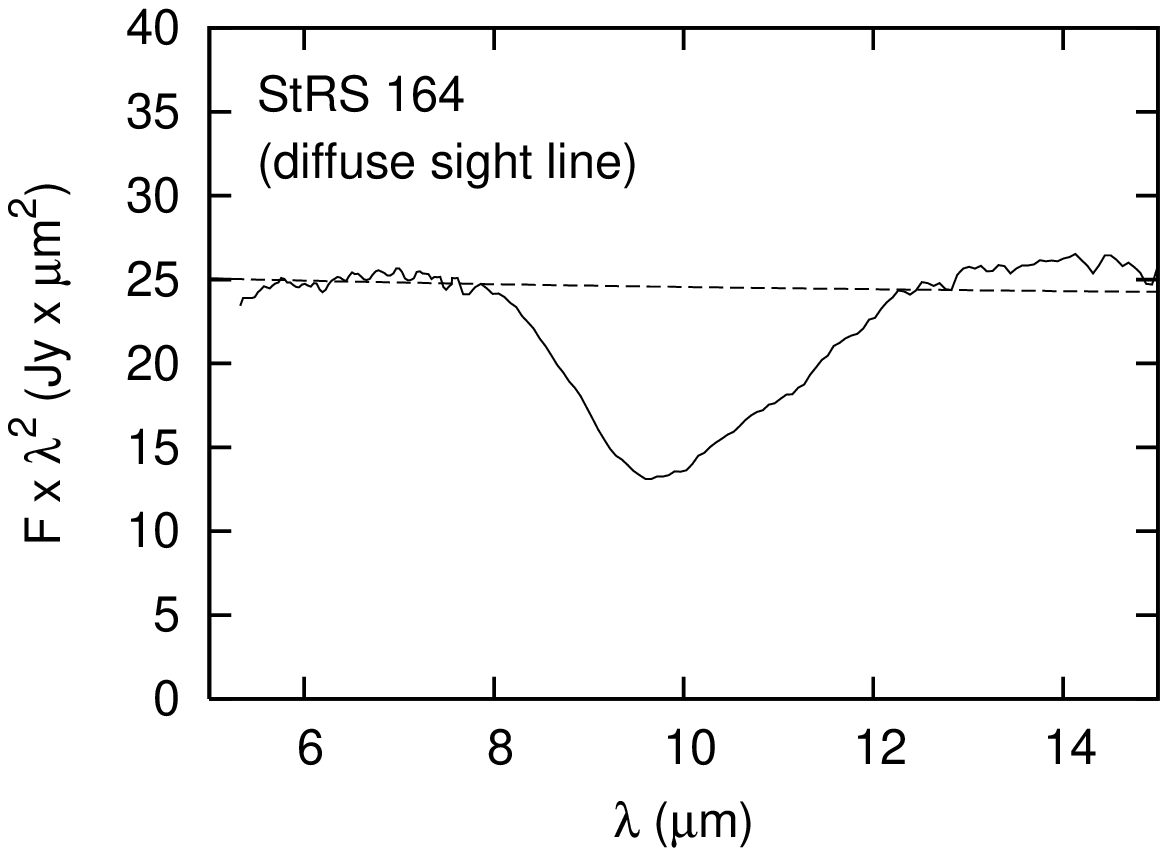}
\includegraphics[angle=270,width=5cm]{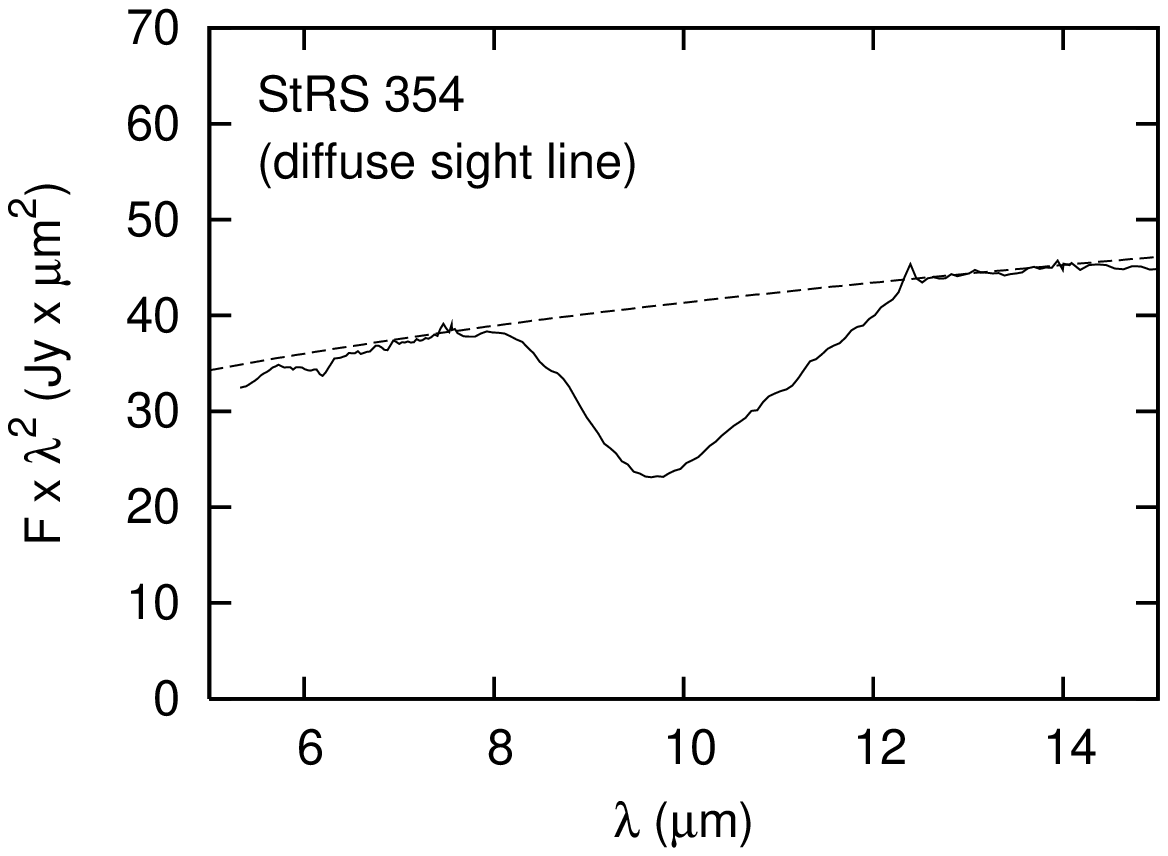}
\includegraphics[angle=270,width=5cm]{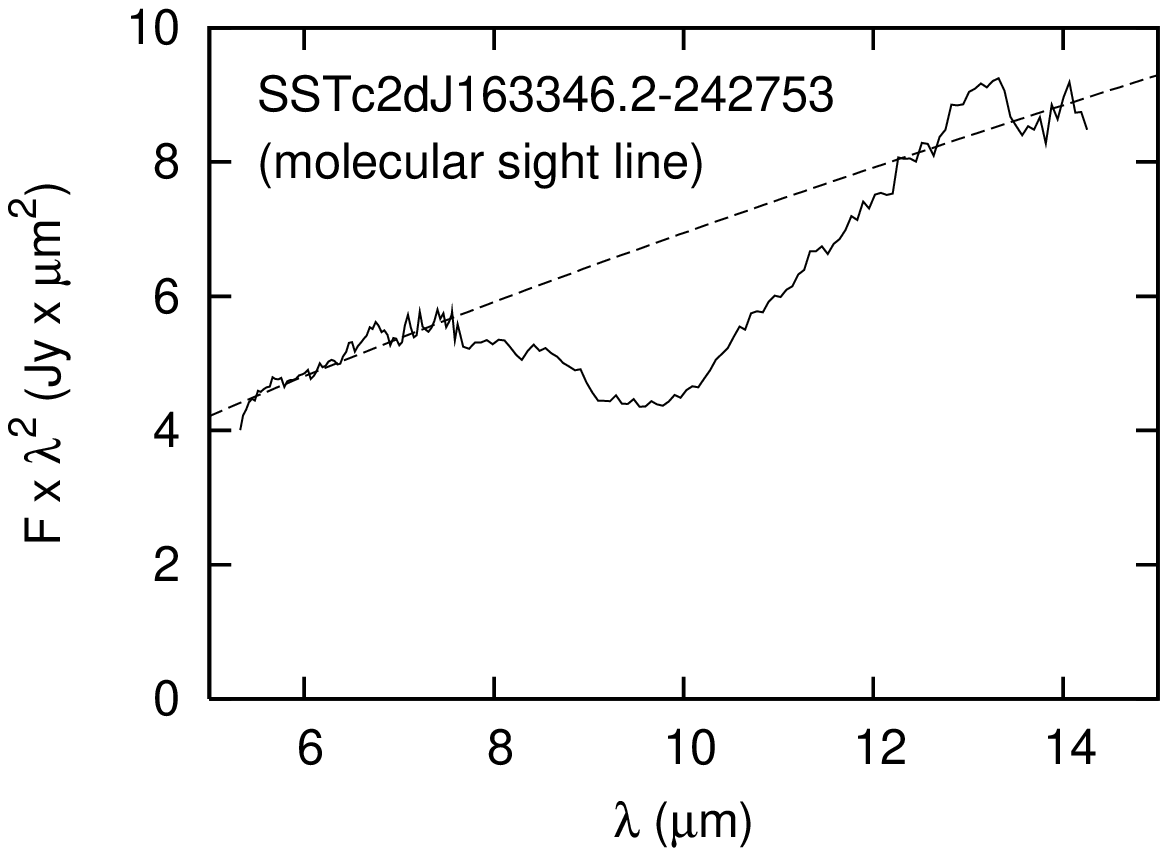}
\includegraphics[angle=270,width=5cm]{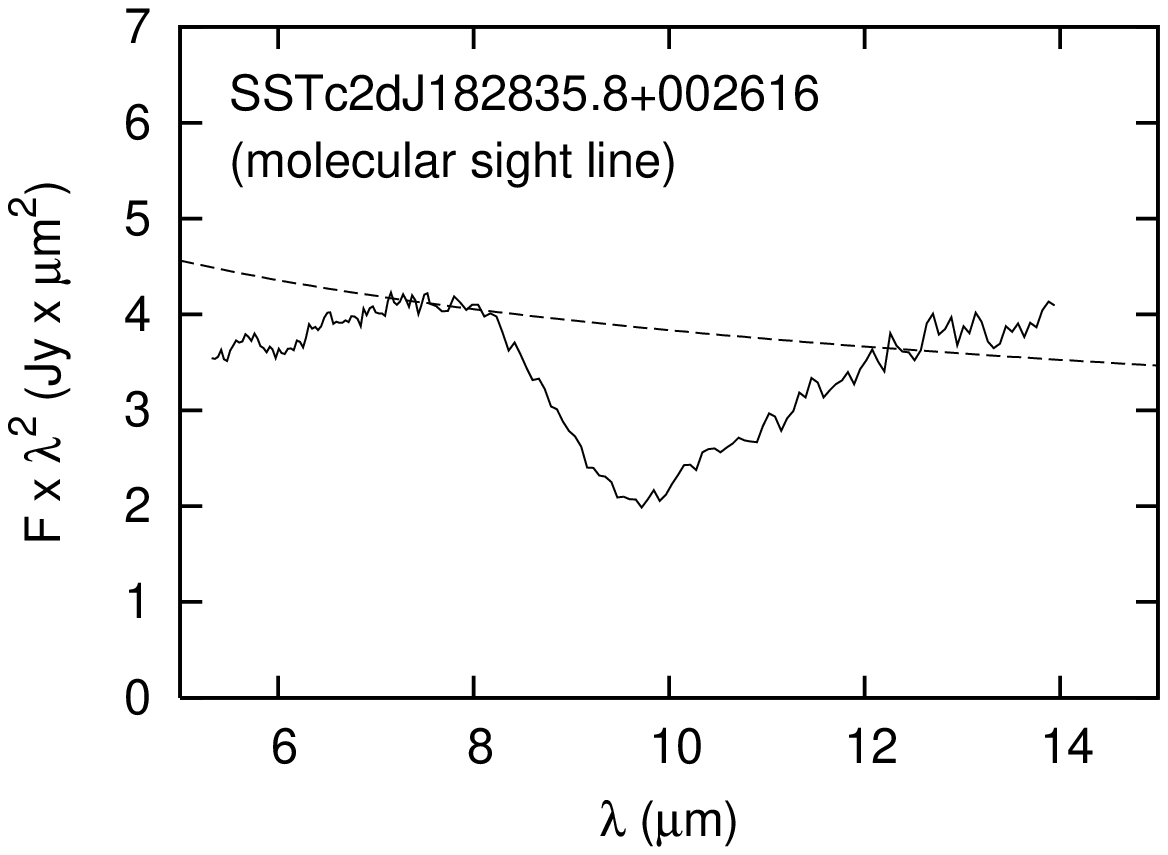}
\includegraphics[angle=270,width=5cm]{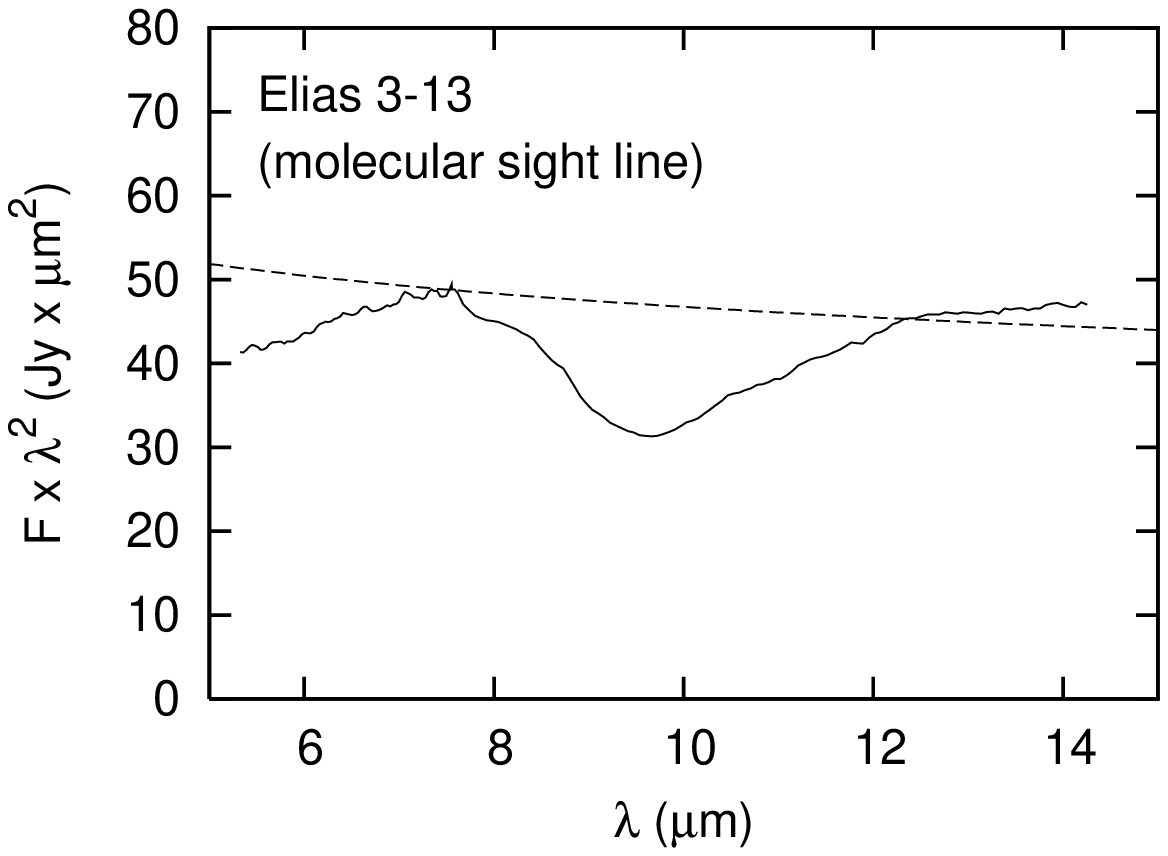}
\includegraphics[angle=270,width=5cm]{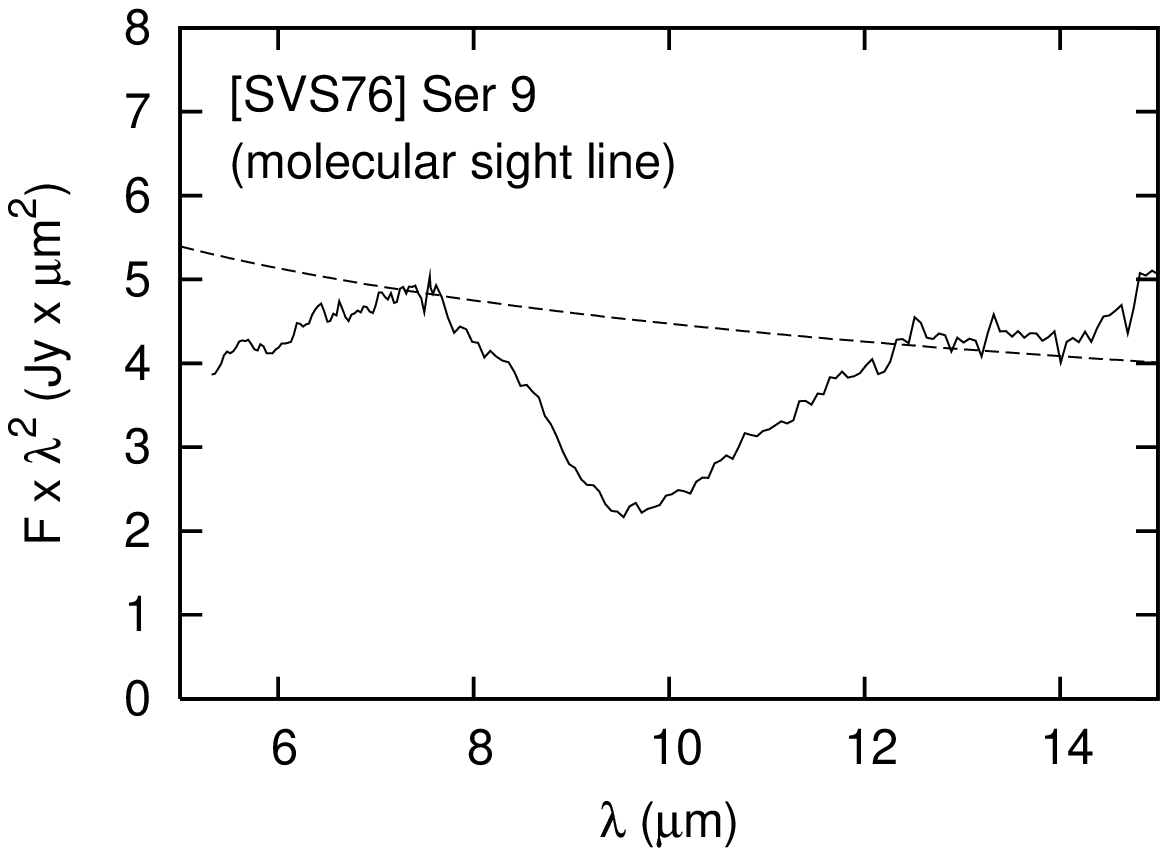}
\caption{Flux multiplied by wavelength squared spectra of the sightlines used for the 9.7 $\mu$m silicate feature study
  together with the continuum {(see main text)}.}
\label{fig:spectra1}
\end{figure*}

\begin{figure*}
\centering
\includegraphics[angle=270,width=5cm]{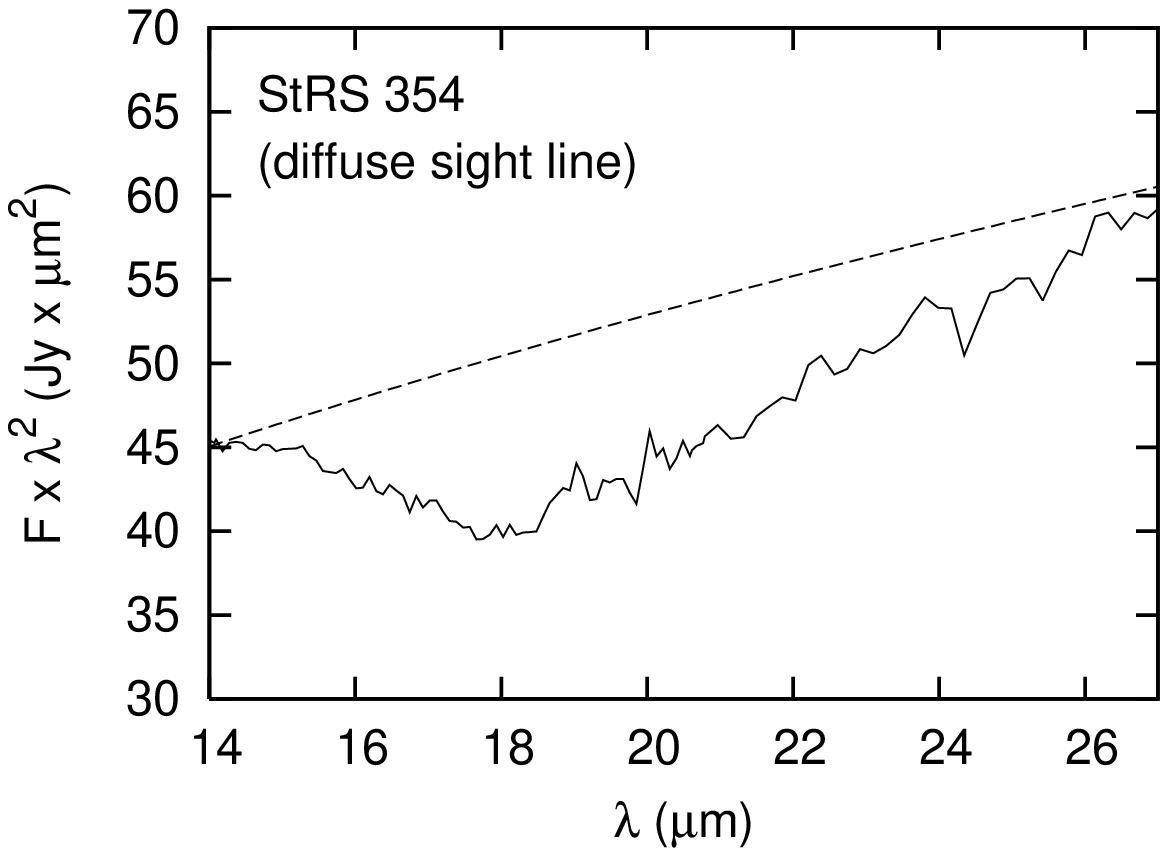}
\includegraphics[angle=270,width=5cm]{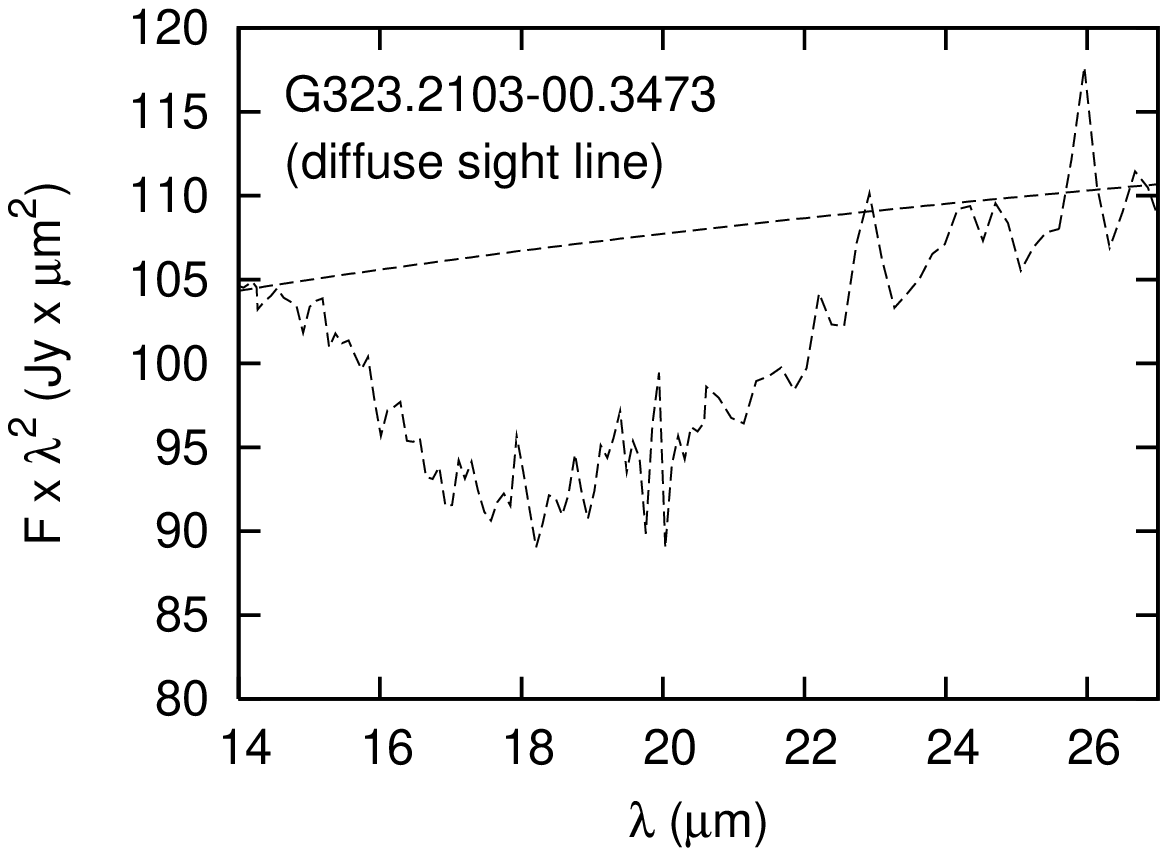}\\
\includegraphics[angle=270,width=5cm]{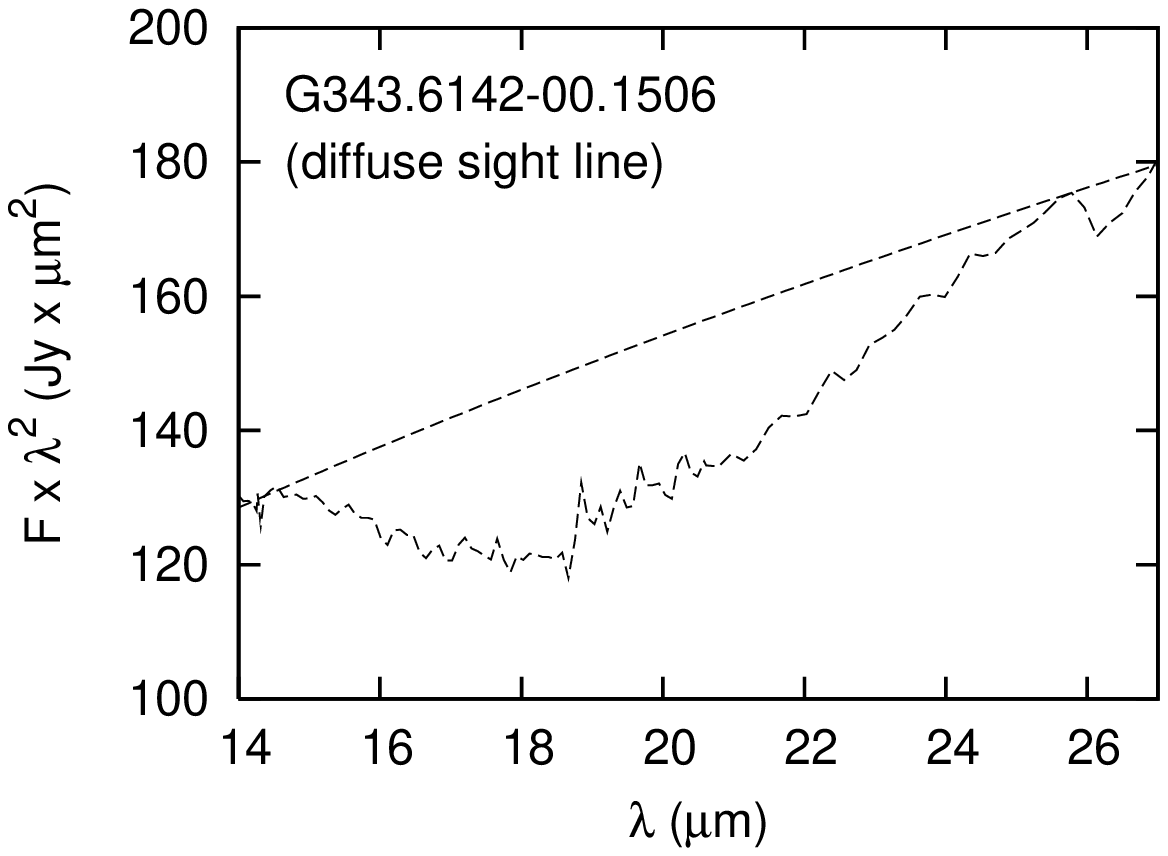}
\includegraphics[angle=270,width=5cm]{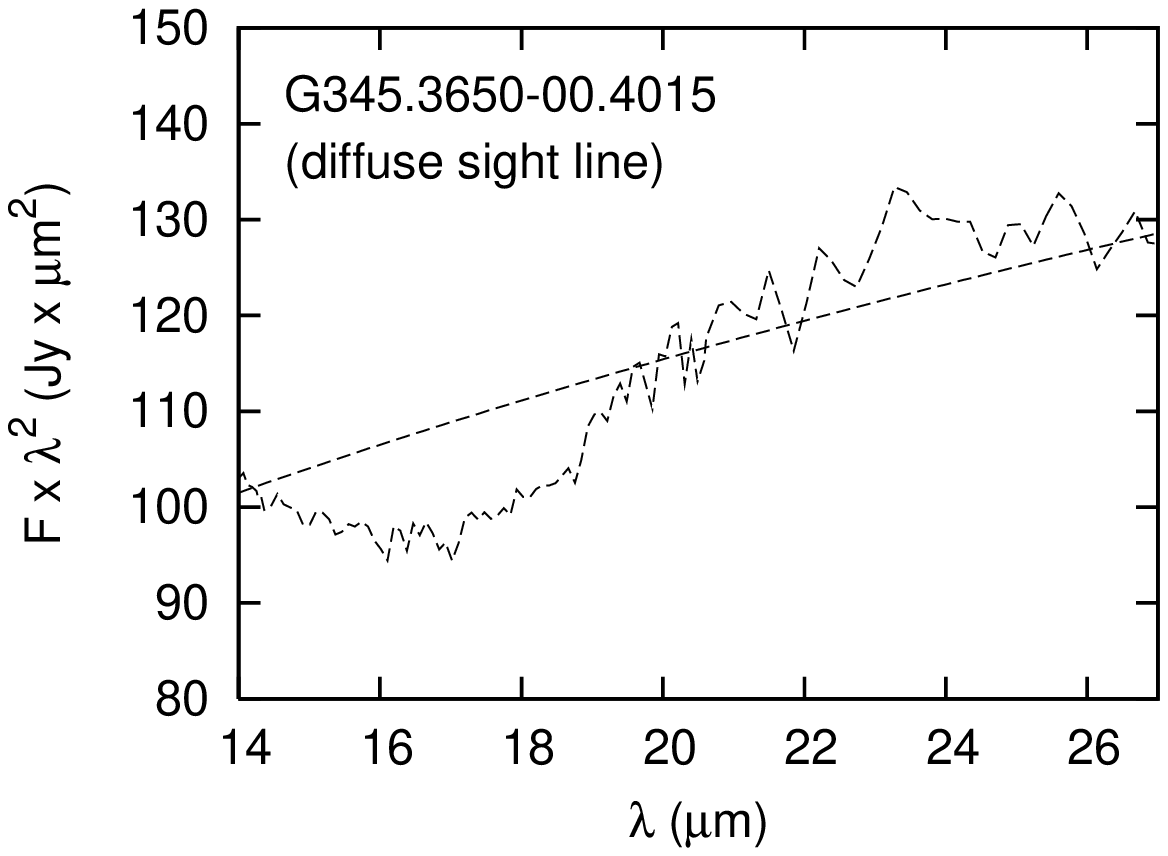}
\caption{Flux multiplied by wavelength squared spectra of the
  sightlines used for the 18 $\mu$m silicate feature study {together
    with the continuum {(see main text)}. Error bars: 2, 3, 3
    and 5 Jy$\times$$\mu$m$^{2}$ for StRS354, G323, G343, and G345
    respectively.}}
\label{fig:spectra2}
\end{figure*}

Basic calibrated data (BCD) {were} obtained from the selected
observations using the Spitzer data reduction pipeline version S15 and
post-BCD 1D spectra were created by carrying out background
subtraction. The resulting spectra are shown in
Fig.~\ref{fig:spectra1} and~\ref{fig:spectra2}.  {We plot these
  as ($F_{\nu} \times \lambda^{2}$) versus $\lambda$ so that an
  unreddened photosphere without photospheric absorption bands shows
  up as a horizontal line, making it easier to recognise the effect of
  extinction on the spectral slope.}

\subsection{The spectra}
 
\begin{figure*}
\centering
\vspace*{1cm}
\includegraphics[angle=270,width=5cm]{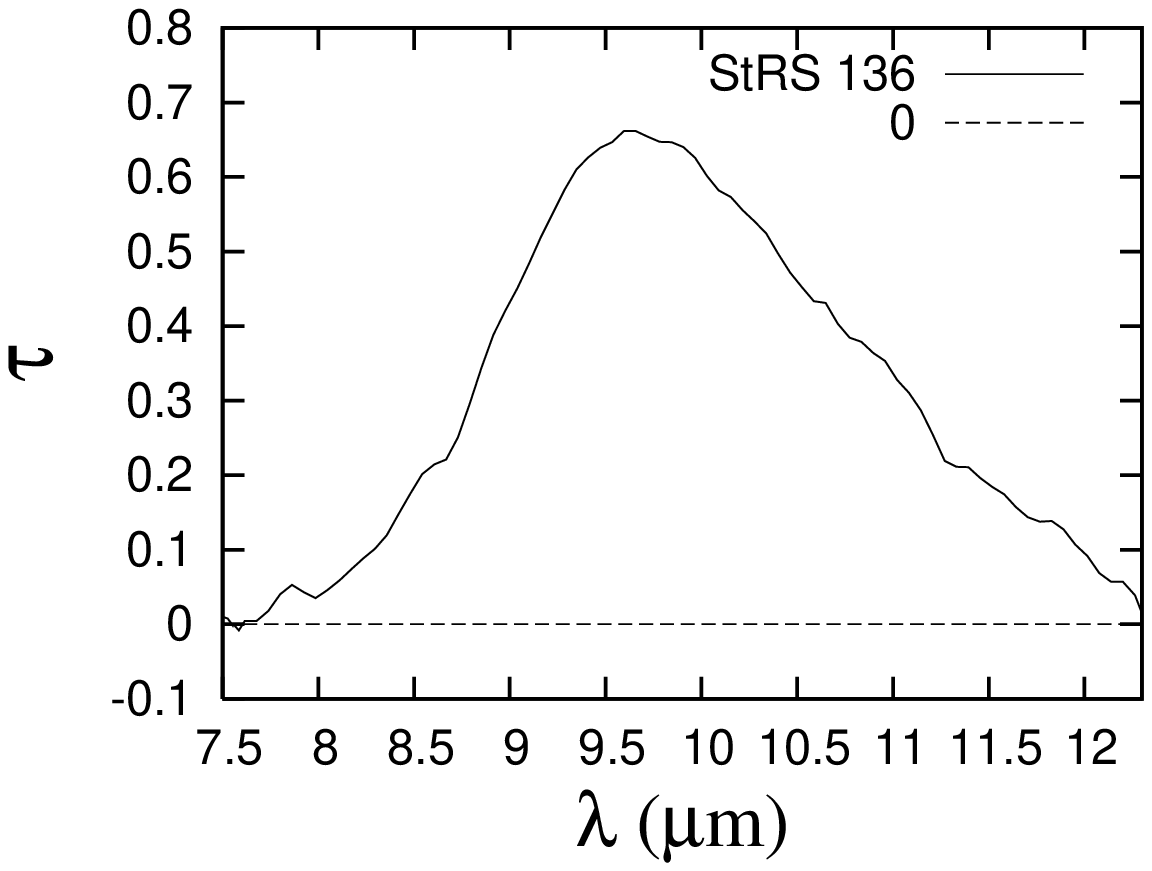}
\includegraphics[angle=270,width=5cm]{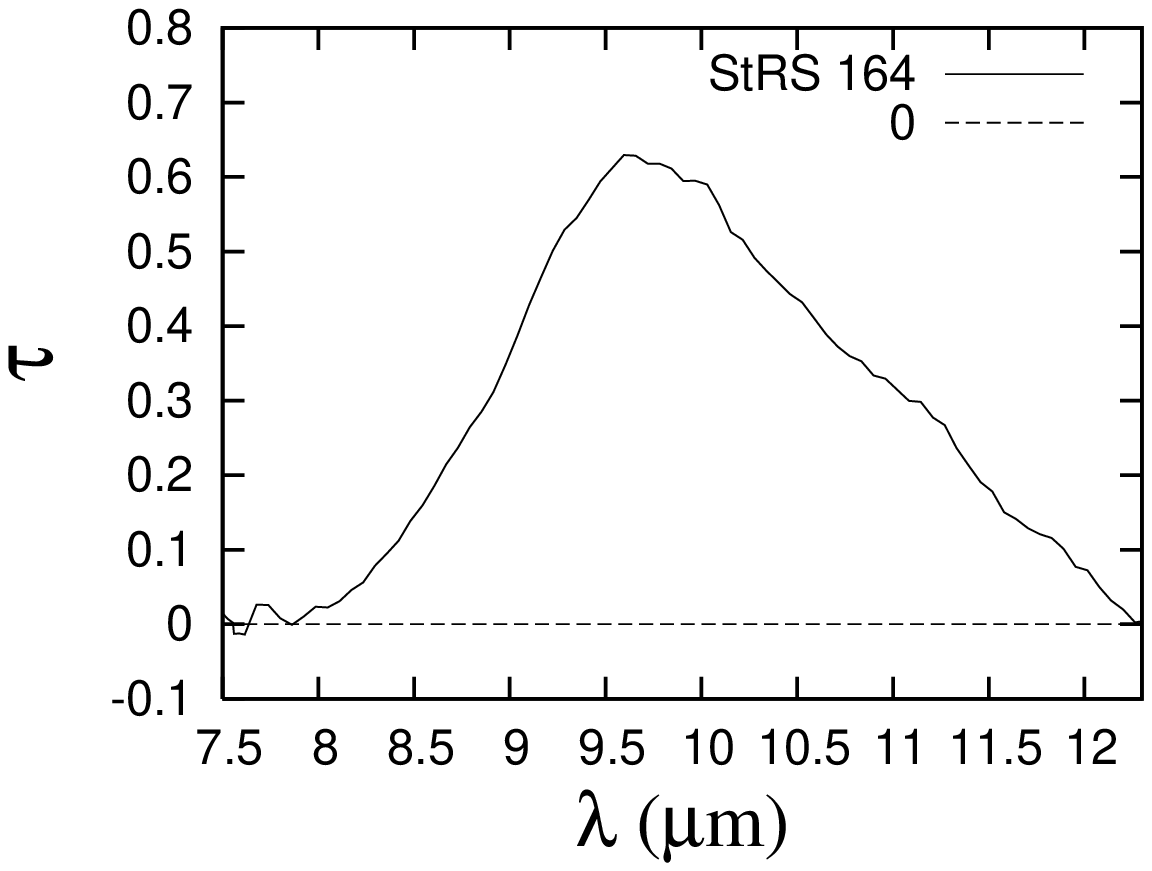}
\includegraphics[angle=270,width=5cm]{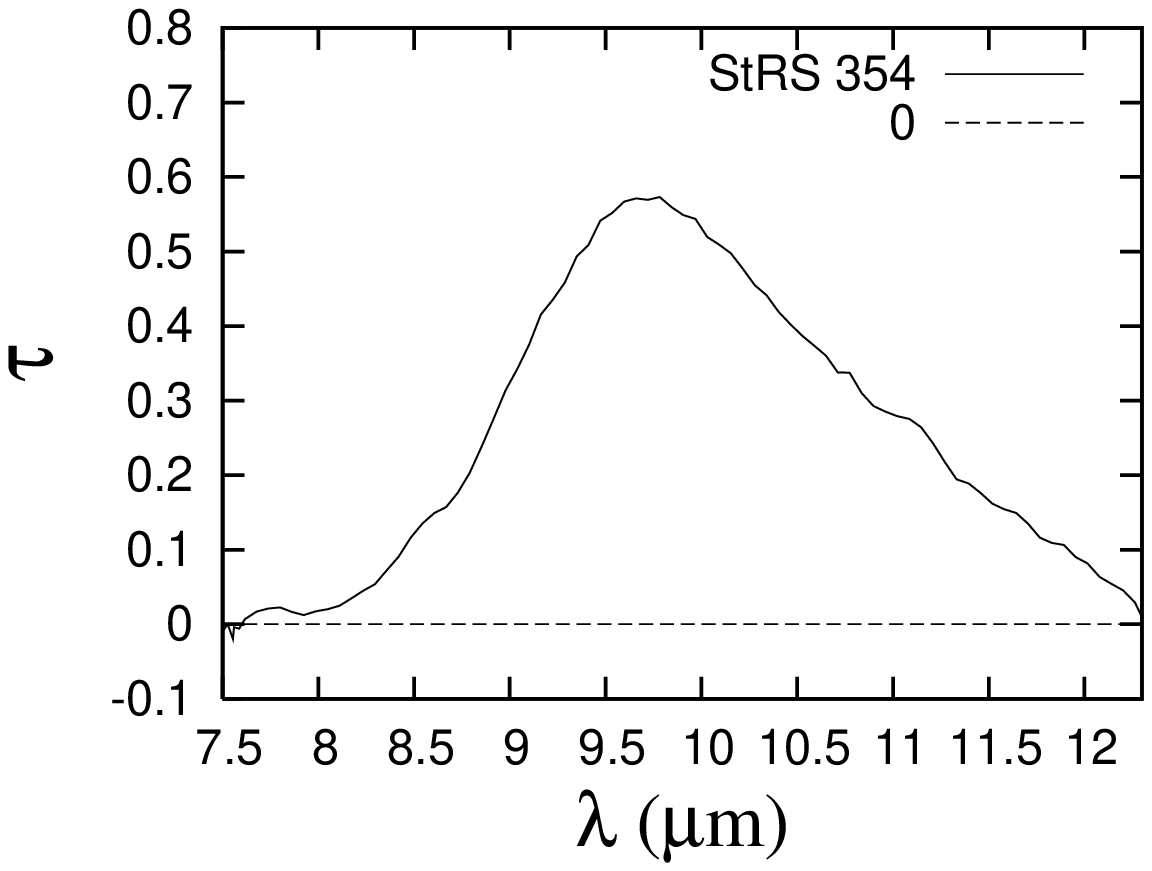}
\includegraphics[angle=270,width=5cm]{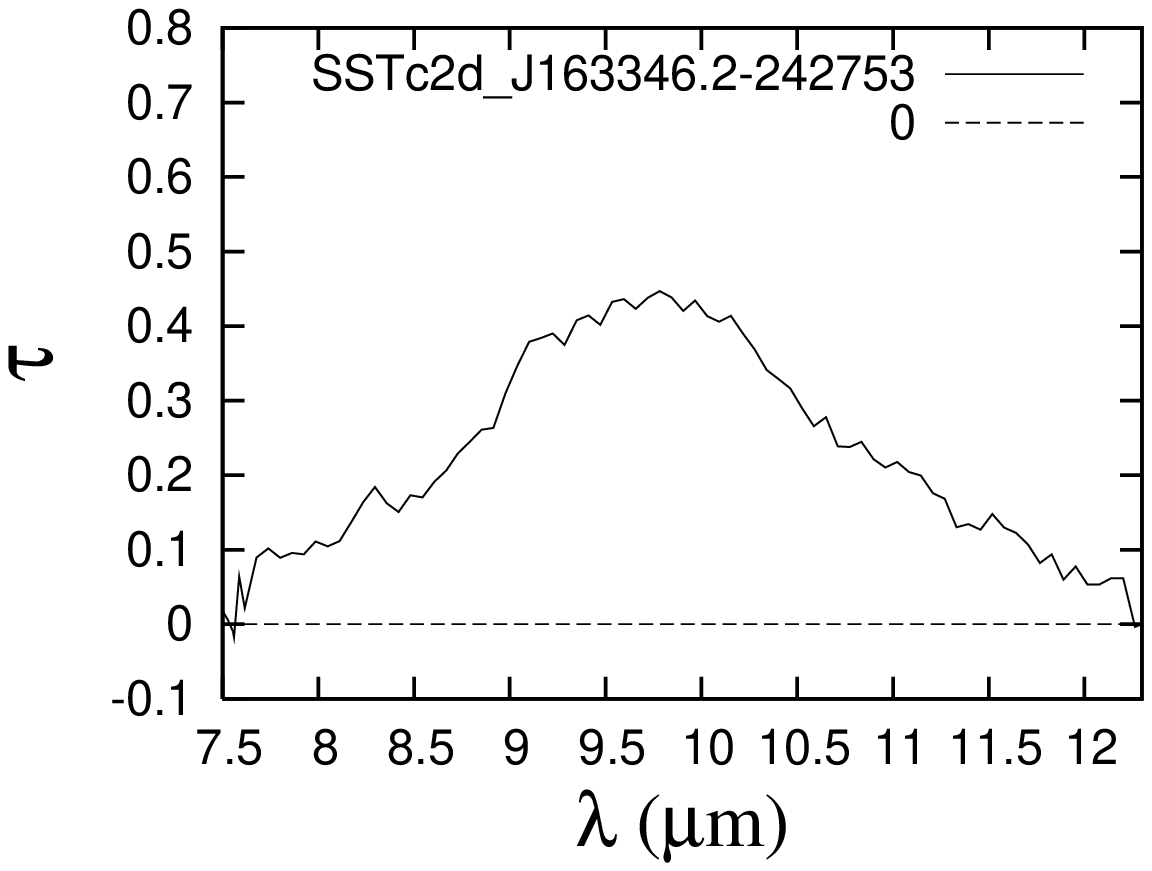}
\includegraphics[angle=270,width=5cm]{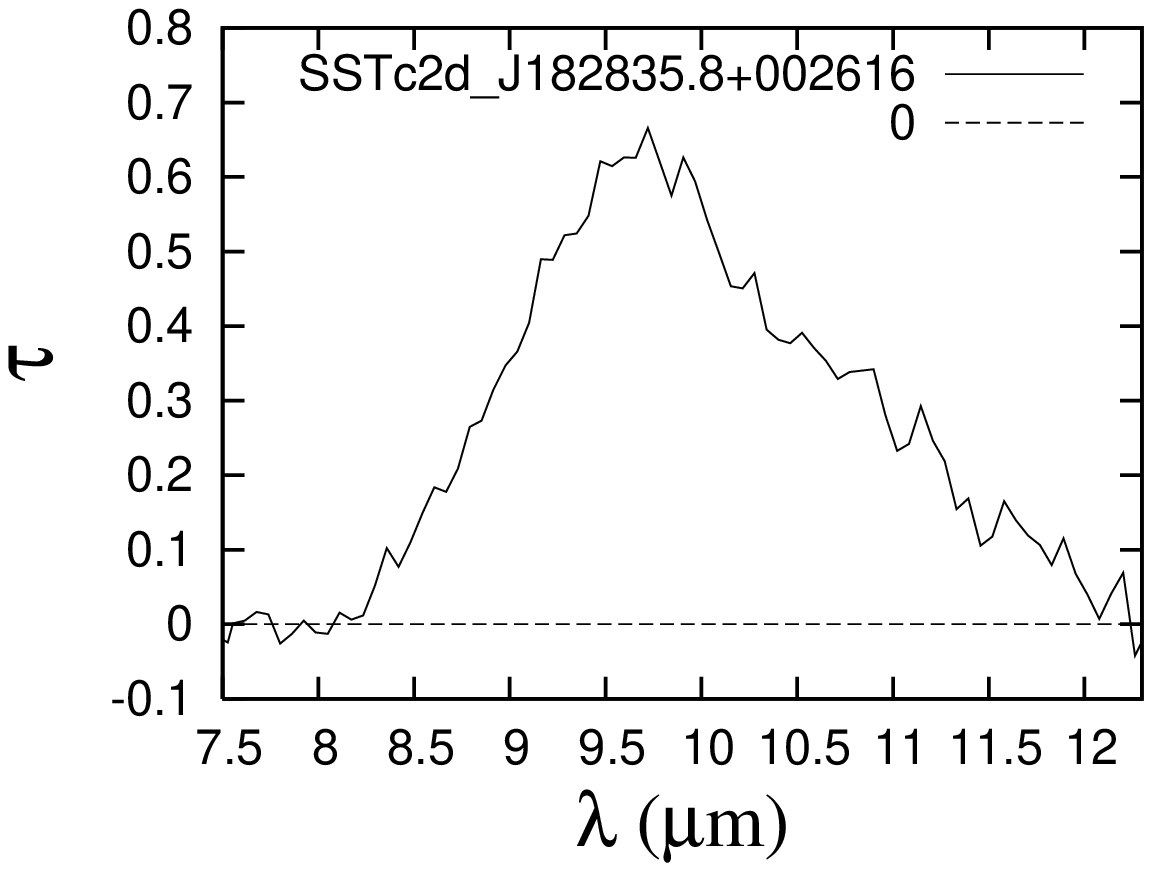}
\includegraphics[angle=270,width=5cm]{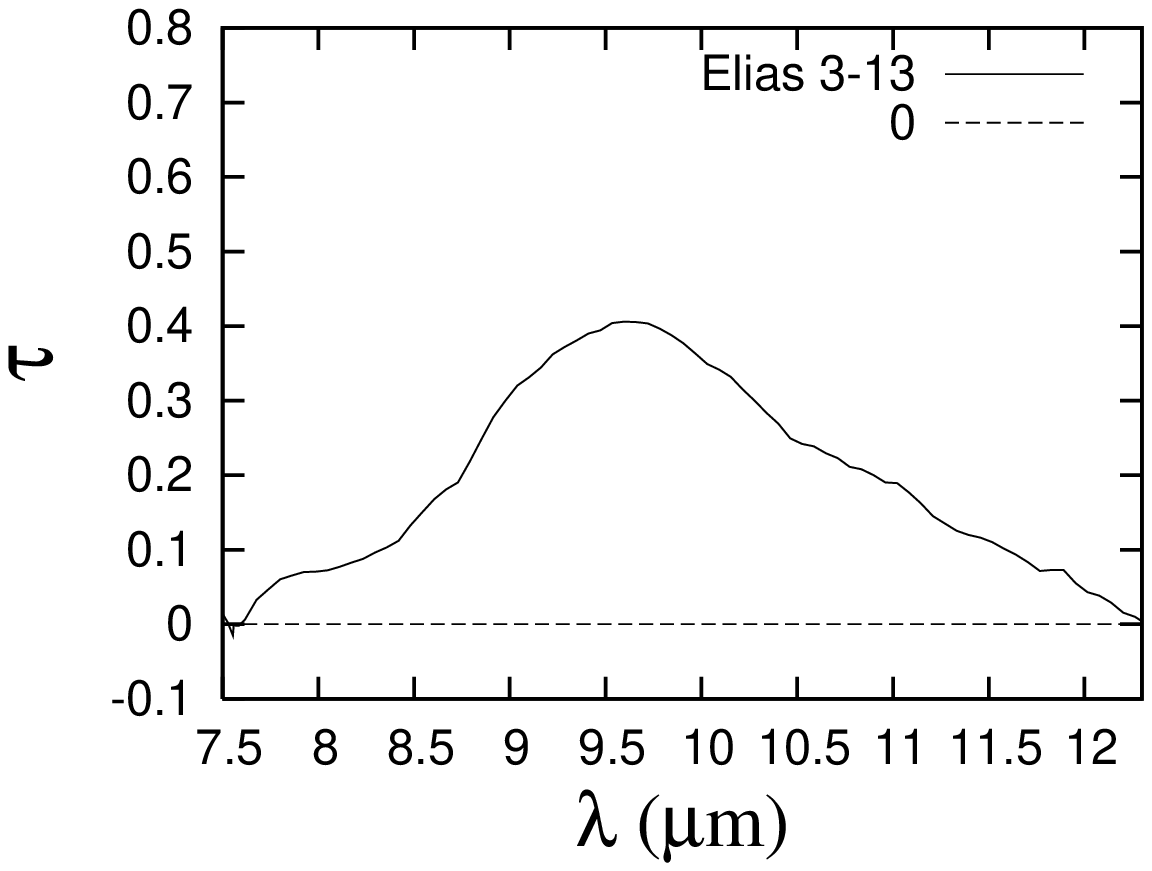}
\includegraphics[angle=270,width=5cm]{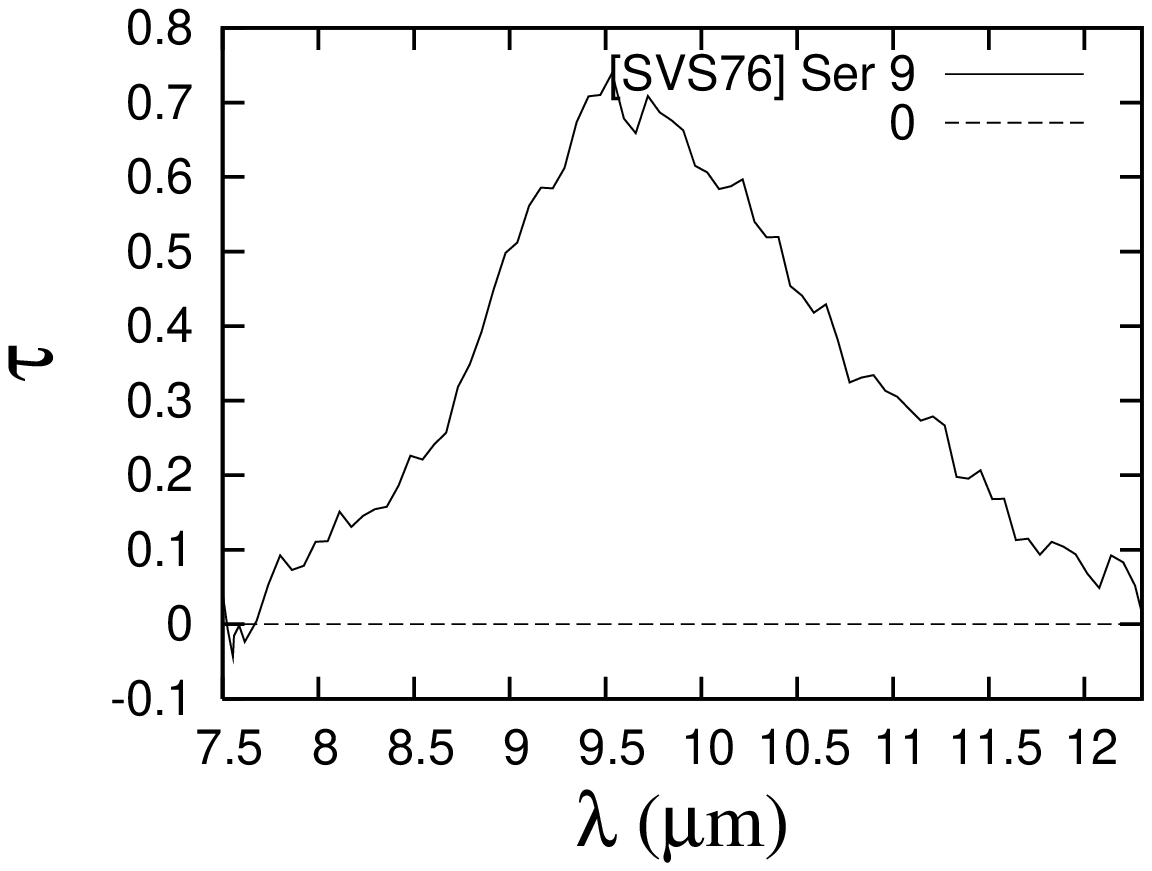}
\caption{The extracted 9.7 $\mu$m optical depth profiles.}
\label{fig:10feats2}
\end{figure*}

{The} 9.7 $\mu$m silicate feature is present in all spectra
{(Figs.~\ref{fig:spectra1} and \ref{fig:10feats2})}. The photospheric
gas phase SiO band is not {apparent} in the spectra, but for only
three sources we can be sure that it is indeed not present, since they
are early type supergiants (see Table~\ref{tab:samplesel}). For the
other sources, however, a very weak SiO band might still be present.

{Ice features are reported to be absent from the Spitzer spectra of
  diffuse sightlines} \citep{Dishoeck04,Boogert04}. Water ice, for
instance, has a feature at about 6 $\mu$m and CH$_{3}$OH ice together
with NH$_{4}^{+}$ ice has a feature at about 6.85 $\mu$m
{\citep[e.g.~][]{2008ApJ...678..985B}}, {which we do not detect in the
  diffuse sightline spectra in this sample.}  This is as expected,
since the conditions in the {diffuse} ISM are not suitable for the
formation of ices. {Ice features are also not detected in the spectra
  towards sightlines with $A_{\mathrm{V}} \lesssim 10$, although the
  6.0 $\mu$m water ice resonance appears only as a weak absorption
  towards sightlines with higher column densities ($A_{\mathrm{V}}
  \gtrsim 10$).}  Therefore, the 9.7 $\mu$m silicate features are
probably also not strongly affected by ice bands, {such as the 12
  $\mu$m water ice librational band}. {Our sightlines do
  probably} not sample the densest parts of the molecular clouds,
where ices are expected to be abundant.

The spectra in Fig.~\ref{fig:spectra2} show the 18 $\mu$m silicate
feature. {In this spectral range the signal to noise level of
  the data is rapidly decreasing due to the lower flux levels. We
  estimate the S/N of the spectra to be in the range 15 to 25. This
  will affect the accuracy of the extracted band profiles.}

\section{Data analysis and results}

\subsection{Extraction of the 9.7 and 18 $\mu$m silicate absorption features}
\label{sect:extraction}

The total extinction by interstellar dust can be divided into a
``continuum'' component (i.e. smoothly varying with wavelength), and
the silicate absorption features, {the shape and strength of which} we
wish to extract. To do this a continuum needs to be specified, which
represents the intrinsic spectrum of the background star reddened by
foreground dust. We assume the intrinsic spectrum of the background
star to be approximately a Rayleigh-Jeans tail, described by $F_{\nu}
\propto \lambda^{-2}$, {while the featureless dust extinction can be
  approximated by a power-law with wavelength as well. The resulting
  continuum, against which the spectral features are seen in
  absorption, is thus represented by a power-law.} This approximation
is good enough for our purposes, since the widths of 9.7 $\mu$m and 18
$\mu$m silicate absorption features each cover a relatively small
wavelength range ($\sim$4 and $\sim$13 $\mu$m, respectively).


The {continuum for the} 9.7 $\mu$m silicate feature is
{defined by the flux levels at} 7.5 {and} 12.3 $\mu$m,
where the first continuum point (at 7.5 $\mu$m) is chosen because
{it does not overlap with potential} photospheric gas phase SiO
bands that might be present in the spectra.  {The resulting slopes
are listed in the last column of Table~\ref{tab:resultstauvsEJK}.}
The {continuum for
  the} 18 $\mu$m silicate feature {is determined separately
  and defined by the values at} 14.2 and 27.0 $\mu$m, {thus
  making sure both continuum points are within the wavelength range
  of} the long wavelength module of Spitzer.
 
\begin{figure*}
\centering
\vspace*{1cm}
\includegraphics[angle=270,width=5cm]{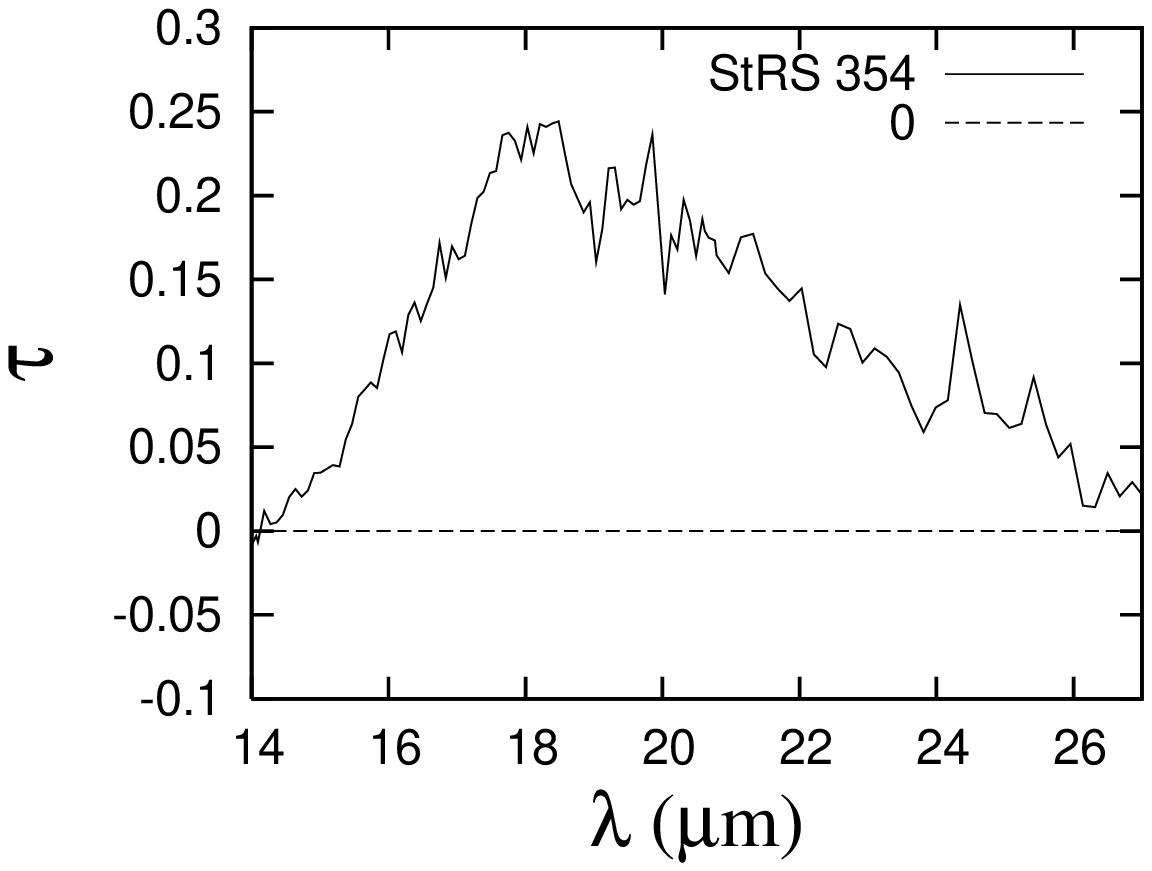}
\includegraphics[angle=270,width=5cm]{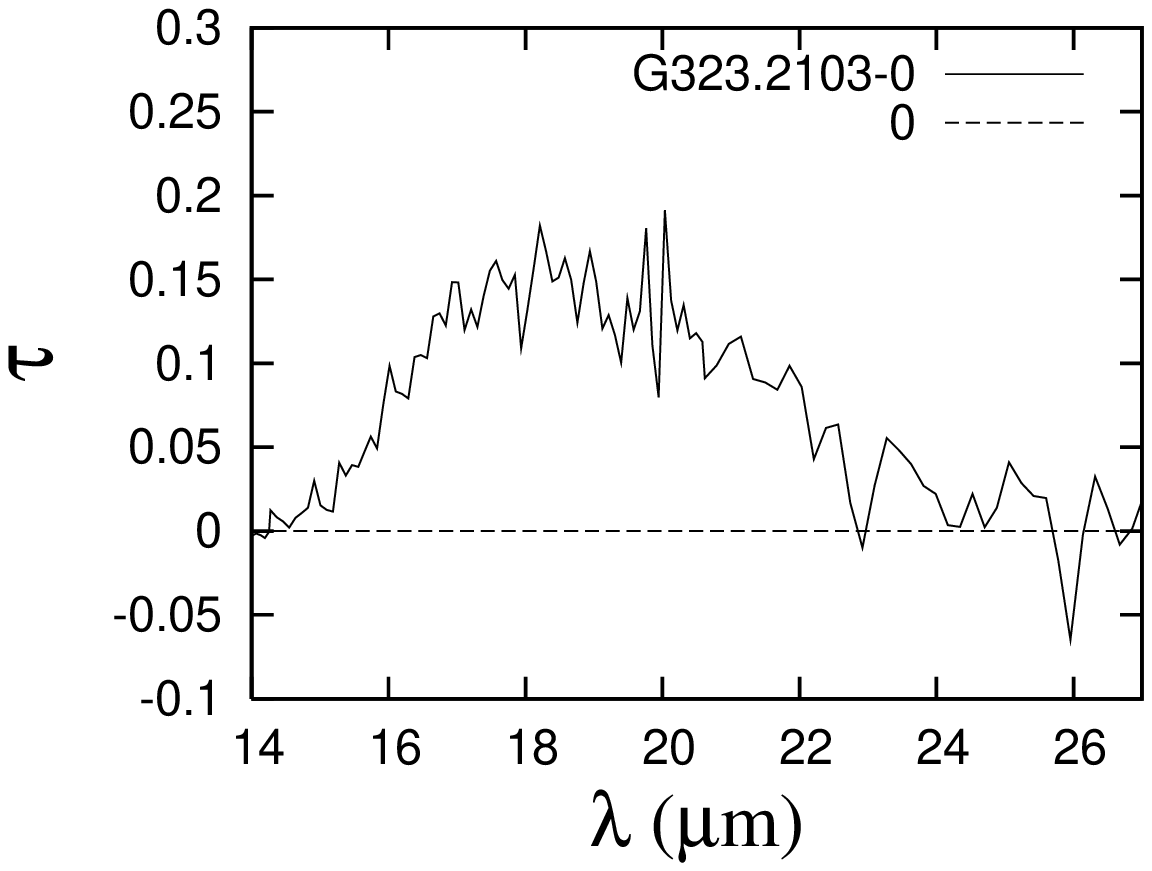}\\
\includegraphics[angle=270,width=5cm]{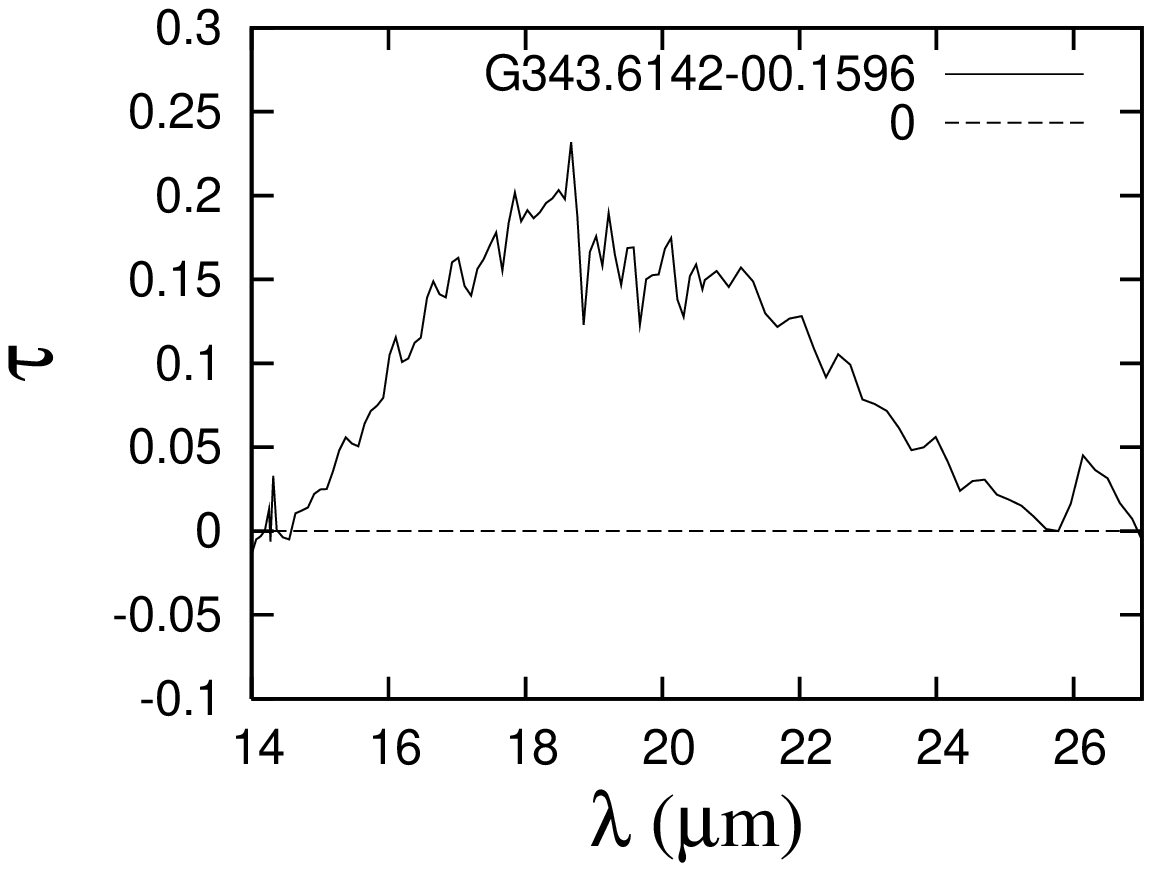}
\includegraphics[angle=270,width=5cm]{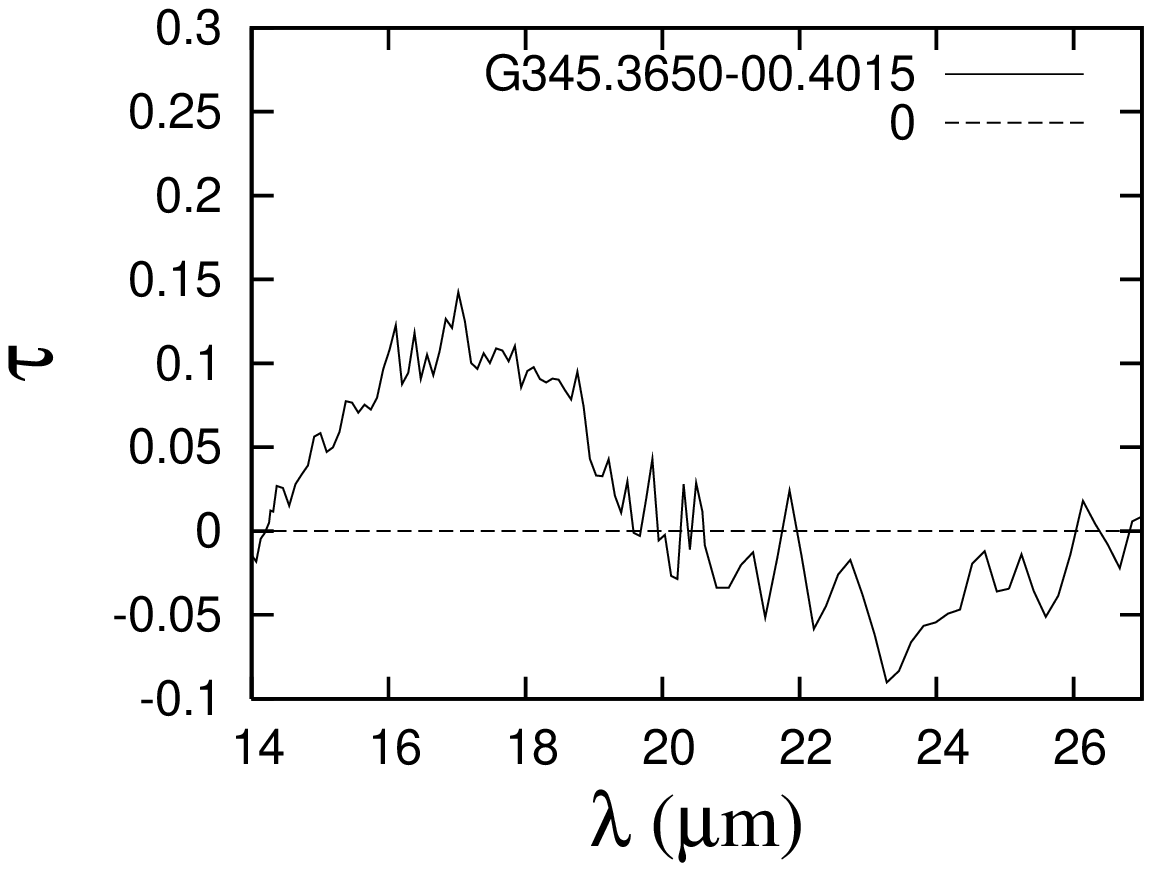}
\caption{The extracted 18 $\mu$m optical depth profiles.}
\label{fig:18feats2}
\end{figure*}

Subsequently, the optical depth profiles are calculated
using:
\begin{equation}
\tau=-\mathrm{log}\,\Bigl( \frac{F_{\nu,\mathrm{obs}}} {F_{ \nu , \mathrm{cont} } } \Bigr)
\end{equation}
where $F_{\nu,\mathrm{obs}}$ is the observed flux and
$F_{\nu,\mathrm{cont}}$ is the flux of the continuum.  The resulting
optical depth profiles are shown in Figs.~\ref{fig:10feats2} and
\ref{fig:18feats2}, {and the peak values for the 9.7 $\mu$m
  feature are also listed in Table~\ref{tab:resultstauvsEJK}. A weak
  correlation exists between $\tau_{9.7}$ and the power-law index of
  the fitted continuum, with a correlation coefficient of $-$0.59.}

\begin{figure}
\centering
\vspace*{1cm}
\includegraphics[angle=270,width=7cm]{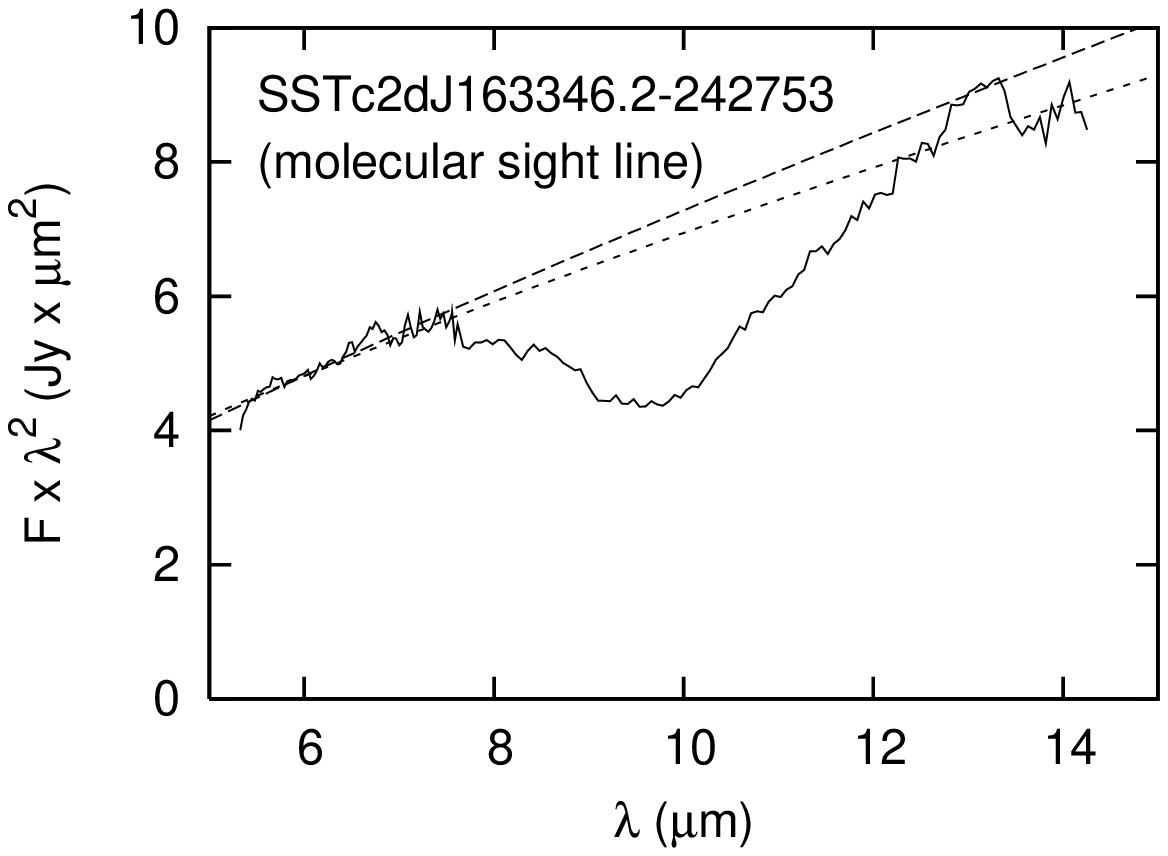}
\includegraphics[angle=270,width=7cm]{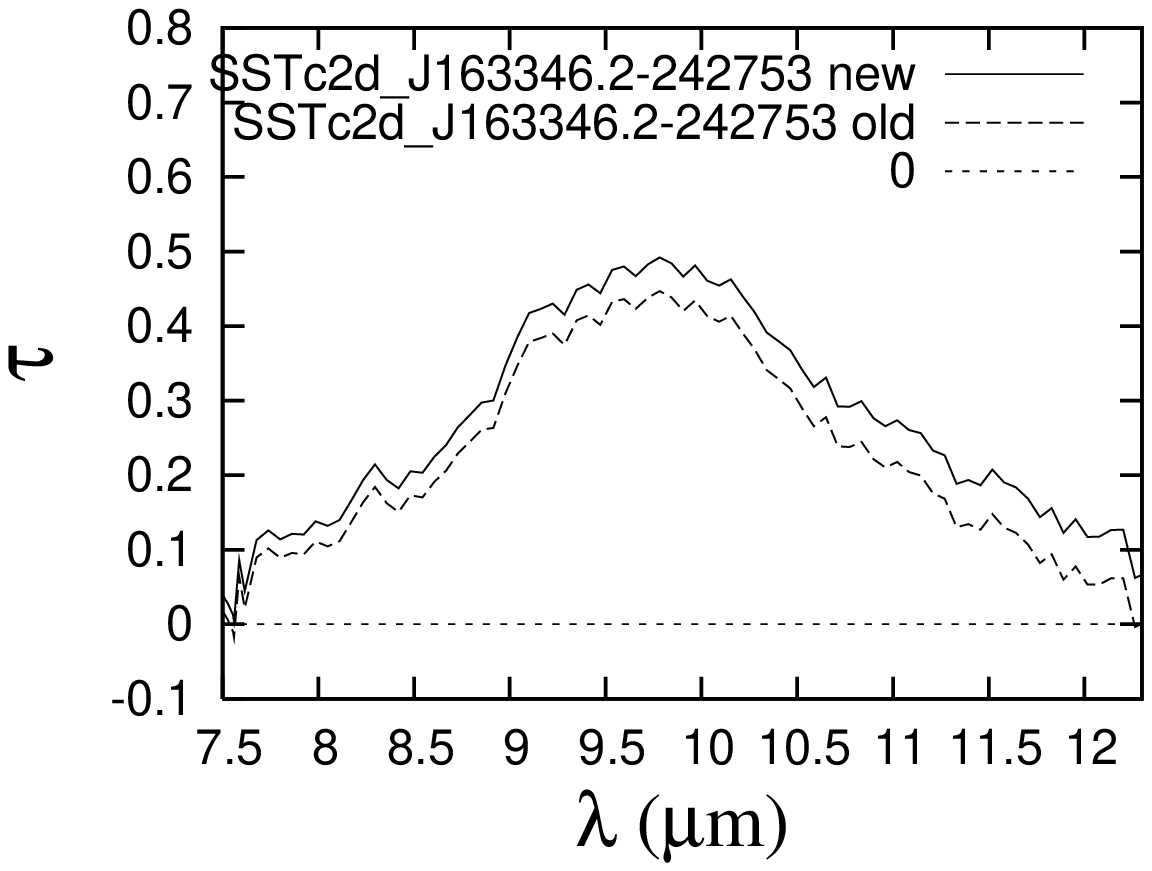} 
\caption{{The effect of choosing a different continuum on the
    shape of the 9.7 $\mu$m silicate band for the molecular sightline
    source SSTc2dJ1633462$-$242753. The rising continuum in the 18
    $\mu$m wavelength range is due to the "continuum extinction" which
    decreases with increasing wavelength.  \textit{upper}: two choices of
    the continuum normalisation; \textit{lower}: comparison of the old
    and new band shape.
}}
\label{fig:f97newcont}
\end{figure}

{The choice of the continuum is important because it affects the shape
  of the extracted band profile and the subsequent analysis. We chose
  to use a systematic approach by always using the same wavelengths at
  which the continuum is defined. We note that the three sources with
  the highest extinction in our sample (all are molecular sightlines)
  show such a sharp drop in flux shortward of 8 $\mu$m that this falls
  below the extrapolated continuum.  We have investigated the impact
  of this choice on our analysis by choosing a different continuum for
  two objects, StRs~164 and SSTc2d\_J163346.2$-$242753 in the 9.7
  $\mu$m region.  The strength of the band changes by at most 10\%,
  when normalised to unity at a wavelength of 10.5 $\mu$m, while the
  shape remains virtually unaffected. This is shown for
  SSTc2d\_J163346.2$-$242753 in Fig.~\ref{fig:f97newcont}. We conclude
  that this different choice of continuum has a minor effect on the
  outcome of our analysis.}

\subsection{$\tau_{9.7}$ versus $E$(J$-$K) relationship}
\citet{Chiar07} investigated the relationship between the strength of
the 9.7 $\mu$m silicate feature, $\tau_{9.7}$, and the {near-infrared}
colour excess, $E$(J$-$K). For diffuse sightlines there is a tight
linear correlation between these two parameters with
$\tau_{9.7}$/$E$(J$-$K)=0.34 \citep{Roche84,Whittet03}, but for the
molecular sightlines this correlation fails
\citep{Chiar07,Whittet88_2}. {The extinction law within the NIR
  wavelength regime appears to remain unchanged with increasing column
  density \citep[e.g.~][]{2007ApJ...664..357R}.}

To compare the sources in our sample to these previous results,
$\tau_{9.7}$ was determined from the 9.7 $\mu$m silicate features in
our sample. An estimation of the error on these measurements was done
by taking different reasonable continua and determining the effect on
$\tau_{9.7}$.

\begin{figure}
\centering
\includegraphics[angle=270,width=7cm]{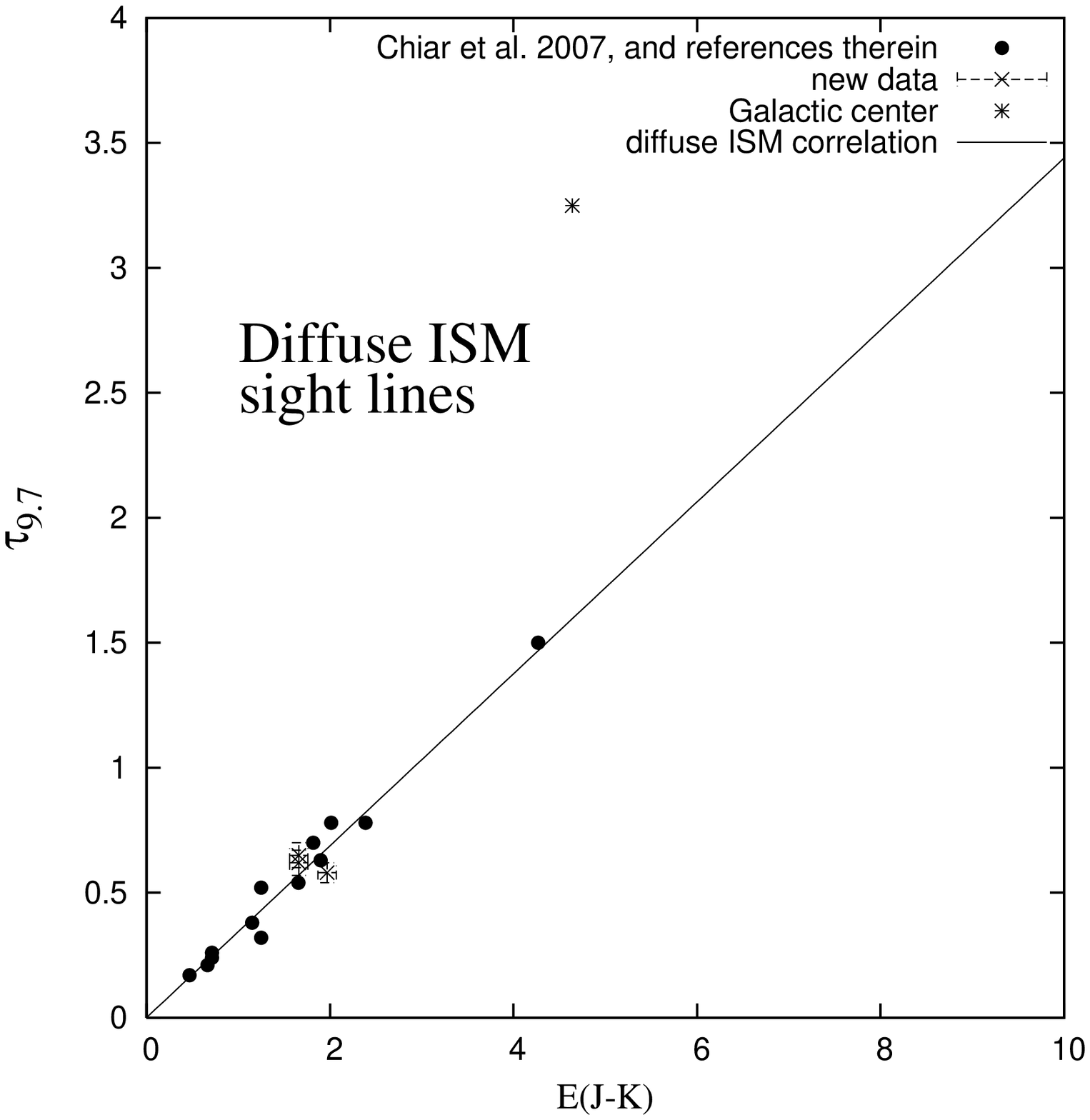}
\includegraphics[angle=270,width=7cm]{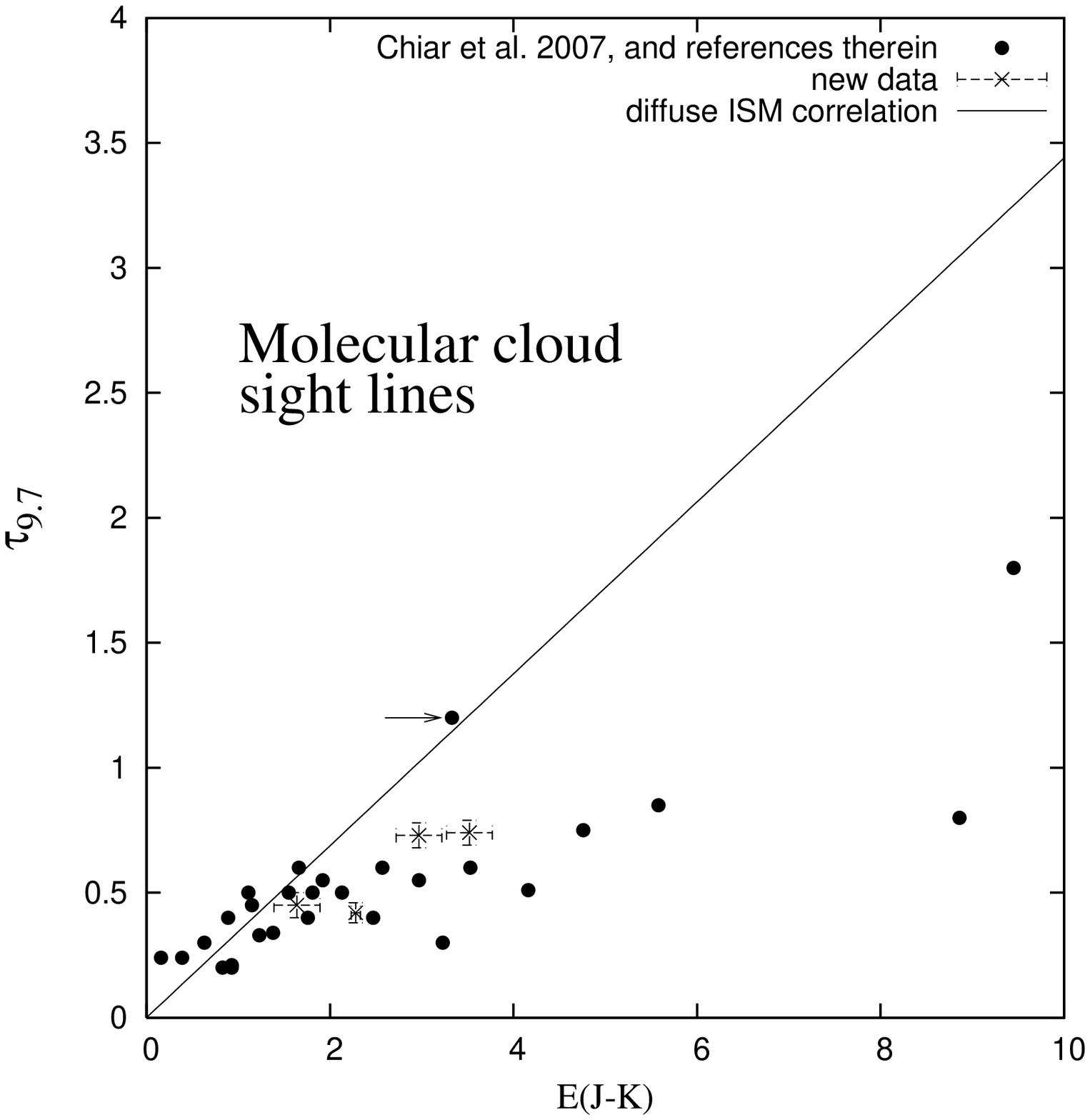}
\caption{The measured optical depth at 9.7 $\mu$m, $\tau_{9.7}$, as a
  function of the J$-$K colour excess, $E$(J$-$K), for diffuse sightlines
  (upper panel) and molecular sightlines (lower panel). The dots
  represent data from \citet[][and references therein]{Chiar07} and
  the crosses represent our measurements. The dashed line is the
  correlation between $\tau_{9.7}$ and $E$(J$-$K) found for the
  diffuse ISM \citep{Chiar07}.The source indicated by the arrow is
  {SSTc2d\_J182852.7+002824}. This source is remarkable, since
  it is the only source with $E$(J$-$K)$>$2 that is still in agreement
  with the diffuse ISM correlation, probably due to the effects of
  ices on the estimated band strength.}
\label{fig:tauvsEJK}
\end{figure}

The $E$(J$-$K) colour excesses were determined by taking the (J$-$K)
colour observed by 2MASS \citep{2MASS06} and subtracting the intrinsic
(J$-$K) colour from \citet{Koornneef83}. If the spectral type of the
source is unknown, we assume it is a G0--M4 giant, with an average
intrinsic (J$-$K) colour of about 0.81 magnitudes (consistent with
\citet{Chiar07}). The errors were estimated by taking into account the
uncertainty in the spectral type and thus the intrinsic (J$-$K)
colour. The results are listed in Table~\ref{tab:resultstauvsEJK} and
Fig.~\ref{fig:tauvsEJK} shows these results together with the Galactic
{Centre} line-of-sight and the results from \citet{Chiar07} for
the diffuse and molecular sightlines separately.

\begin{table*}
\centering
\begin{tabular}{llllllll}
\hline \\
target name &  Spectral type & Diffuse/Molecular & (J$-$K)$_{int}$$^{1}$ & (J$-$K)$_{obs}$ & $E$(J$-$K) & $\tau_{9.7}$ & {power-law index}\\
\hline\\
StRS 136 & B8--A9I$^{2}$ & D & 0.12 & 1.78 & 1.66$\pm$0.1 & 0.65$\pm$0.05 & {$-$2.00} \\
StRS 164 & B8--A9I$^{2}$ & D & 0.12 & 1.78 & 1.66$\pm$0.1 & 0.62$\pm$0.05 & {$-$2.03}\\
{SSTc2d\_J182835.8+002616} & \ldots & M (Serpens) & 0.81 & 3.78 & 2.97$\pm$0.40 & 0.73$\pm$0.05 &{$-$2.25}\\
{$[$SVS76$]$ Ser 9} & \ldots & M (Serpens) & 0.81 & 4.33 & 3.52$\pm$0.40 & 0.74$\pm$0.05 & {$-$2.27}\\
StRS 354 & O7--B3$^{2}$ & D & $-$0.09 & 1.88 & 1.97$\pm$0.1 & 0.58$\pm$0.04&{$-$1.73}\\
{Elias 3-13} & K2III$^{3}$ & M (Taurus) & 0.73 & 3.01 & 2.28$\pm$0.05 & 0.42$\pm$0.04 &{$-$2.15}\\
SSTc2d\_J163346.2$-$242753 & \ldots & M ($\rho$ Ophiuchi) & 0.81 & 2.45 & 1.64$\pm$0.25 & 0.45$\pm$0.05 &{$-$1.28}\\
\hline\\
\end{tabular}
\caption{The results for $\tau_{9.7}$ and $E$(J$-$K) measured for the sources in our sample. The target name, spectral type (if known) and the type of sightline (diffuse, D, or molecular, M) are listed. If the spectral type is unknown we assume it is a G0--M4 giant. The corresponding intrinsic J$-$K colour, (J$-$K)$_{int}$, and the J$-$K colour as observed by 2MASS, (J$-$K)$_{obs}$, are also listed. The difference between these two values gives the colour excess $E$(J$-$K). {The last two columns list the} the measured optical depth at 9.7 $\mu$m, $\tau_{9.7}$, {and the index of the power-law used to determine the continuum around 10 $\mu$m}. $^{1}$\citet{Koornneef83}, $^{2}$\citet{Rawlings00}, $^{3}$\citet{Elias78})}
\label{tab:resultstauvsEJK}
\end{table*}

Our measurements are in agreement with the results found by
\citet{Chiar07} as can be seen in Fig.~\ref{fig:tauvsEJK}. Please
note, that two of our measurements are for {for sightlines also
  in the sample of} \citet{Chiar07}, namely: {Elias 3-13} and
{[SVS76] Ser 9} and we have replaced their measurements by ours
in Fig.~\ref{fig:tauvsEJK}. The two measurements for {[SVS76]
  Ser 9} agree within the uncertainty. For {Elias 3-13} the
measurement was done by~\citet{Whittet88_2} from a relatively noisy,
ground based spectrum, so the error on their measured $\tau_{9.7}$ is
quite large (approximately 0.08 in $\tau$). Moreover, they did not
correct for photospheric gas phase SiO.

We also included the Galactic {Centre} line-of-sight in
Fig.~\ref{fig:tauvsEJK} (upper panel). For this we used
$\tau_{9.7}$=3.25 \citep{Kemper04, Min07} and
$\tau_{9.7}$/$E$(J$-$K)=0.7 \citep{Roche85}. The Galactic {Centre}
line-of-sight is exceptional, since it is the only one for which
$\tau_{9.7}$ versus $E$(J$-$K) lies above the diffuse ISM correlation.
This sightline probes the local ISM as well as the inner parts of our
Galaxy in contrast to the other sightlines which only probe the local
ISM. Since there are relatively fewer C-rich stars (as compared to
O-rich stars) around the Galactic {Centre} than in the outer regions
of our Galaxy, the {relative} abundance of {carbon-rich}
dust is probably also lower around the Galactic {Centre}
\citep{Roche85}. This would explain the relatively high
$\tau_{9.7}$/$E$(J$-$K) ratio for this particular line-of-sight, since
carbon rich dust {is a major component of} the {near-infrared}
extinction.

\begin{figure}
\centering
\includegraphics[angle=270,width=7cm]{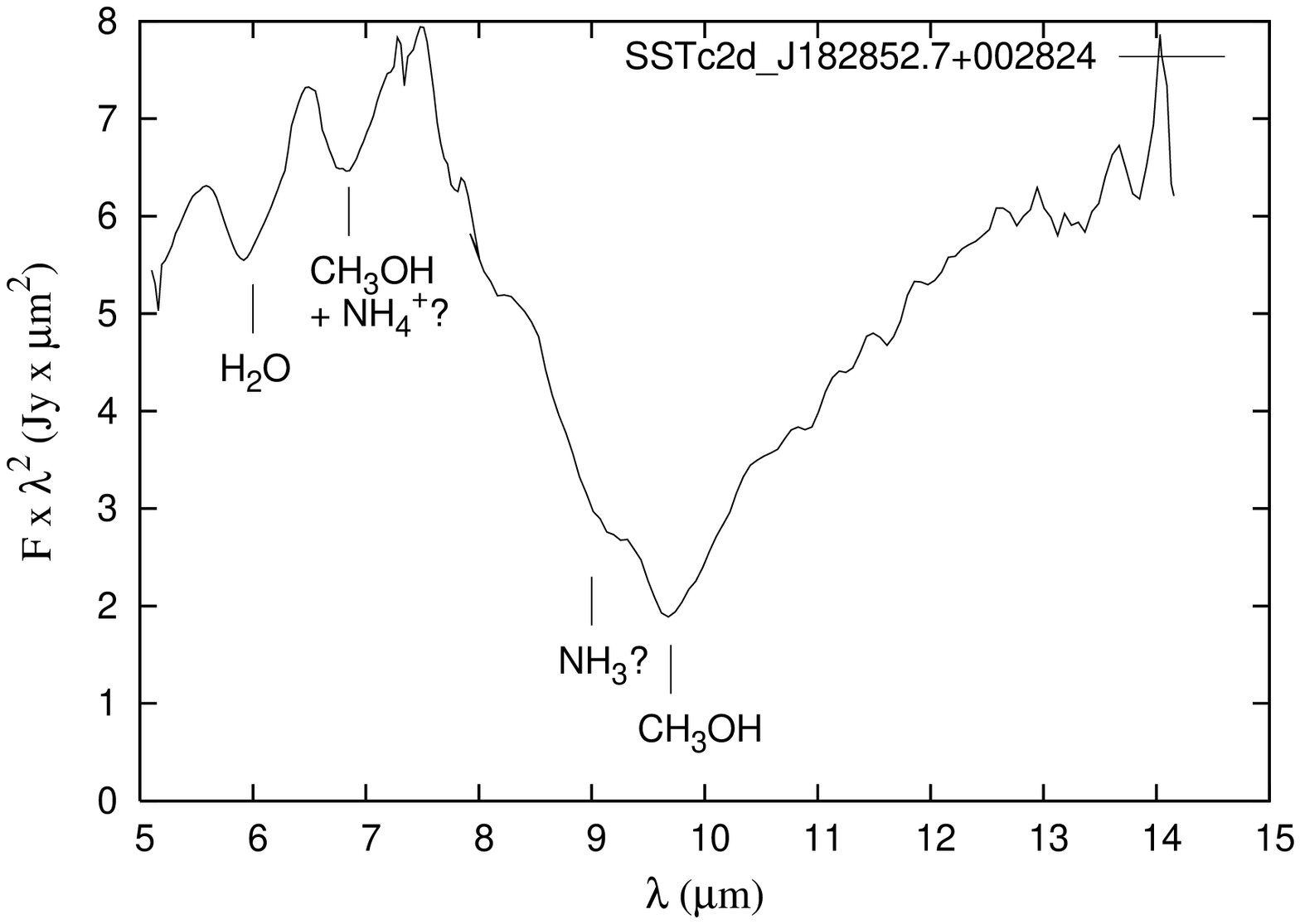}
\caption{The spectrum observed towards
  {SSTc2d\_J182852.7+002824}. Absorption features caused by
  ices on top of the silicate feature are indicated. The absorption
  feature caused by CH$_{3}$OH ice at about 9.7 $\mu$m makes the
  silicate feature seem deeper than it really is. Consequently, the
  strength of the 9.7 $\mu$m silicate feature is
  overestimated. }
\label{fig:sst18}
\end{figure}

Furthermore, there is one molecular cloud source (indicated by the
arrow in Fig.~\ref{fig:tauvsEJK}) that seems to agree very well with
the diffuse ISM correlation despite the high value for $E$(J$-$K),
{while other molecular sightlines with $E$(J$-$K) $\gtrsim 2$
  deviate from this correlation. This is SSTc2d\_J182852.7+002824,
  which} was observed with Spitzer in the c2d legacy program
{\citep{2003PASP..115..965E}. Its reduced} spectrum is shown in
Fig.~\ref{fig:sst18}. Prominent ice features are present at about 6
and 6.85 $\mu$m caused by H$_{2}$O and CH$_{3}$OH + NH$_{4}^{+}$ ice
respectively \citep{Boogert04,Dishoeck04}. The latter also has a
feature at about 9.7 $\mu$m, which increases the depth of the 9.7
$\mu$m silicate feature {leading to an overestimate of the
  absorption due to silicates. Therefore, \citet{Chiar07} {apparently}
  overestimated the absorption due to silicates. Correcting for ice
  absorption would move the $\tau_{9.7}$ point downward in
  Fig.~\ref{fig:tauvsEJK}, which is in agreement with the results for
  the other dense cloud sources.}  We note the large differences in
the strength of the ice bands between the spectrum in
Fig.~\ref{fig:sst18} and the 4 molecular cloud sightlines shown in
Fig.~\ref{fig:spectra1}, for similar values of
$E$(J$-$K). {Apparently, the observed changes in the
  $\tau_{9.7}$ versus $E$(J$-$K) relationship are not related to ice
  formation.}

\subsection{Comparison of the 9.7 {$\mu$m} silicate profiles}
The shape of the 9.7 $\mu$m silicate absorption features in our sample
(see Fig.~\ref{fig:10feats2}) will be compared and discussed for the
diffuse and molecular sightlines separately below. Throughout this
article we will use the feature determined for the Galactic
{Centre} by \citet{Kemper04} as a reference.

\subsubsection{Diffuse sightlines}
\begin{figure}
\centering
\includegraphics[angle=270,width=8cm]{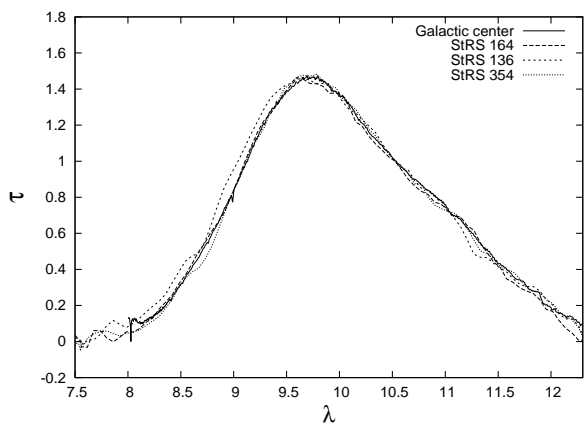}
\caption{The 9.7 $\mu$m silicate features in terms of optical depth
  for the diffuse sightlines together with the feature observed
  towards the {Galactic Centre} \citep{Kemper04}. The features
  are normalised to unity at 10.5 $\mu$m.}
\label{fig:earlydiff}
\end{figure}

Figure~\ref{fig:earlydiff} shows the extracted features for the
diffuse sightlines in our sample together with the Galactic
{Centre} feature from \citet{Kemper04}. All features are
normalised to unity at 10.5 $\mu$m, because the contribution of any
photospheric gas phase SiO bands that might be present in the spectra
is negligible from about 10.5 $\mu$m onwards. This {effect} is
not important for the diffuse sightlines, since the background sources
are known to be early type supergiants without photospheric gas phase
SiO, but it will matter for the molecular sightlines which we will
discuss later.

\begin{figure}
\centering
\includegraphics[angle=270,width=8cm]{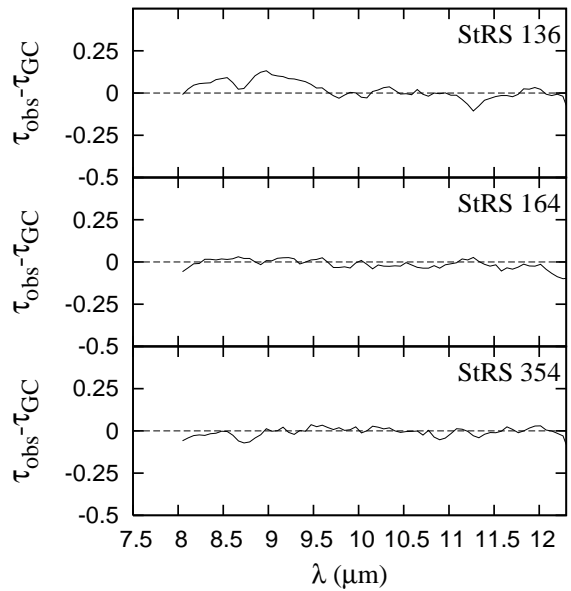}
\caption{The difference between the observed features in the diffuse
  sightlines in our sample and the {Galactic Centre} feature
  derived by \citet{Kemper04}.}
\label{fig:StRS_galcent}
\end{figure}

From Fig.~\ref{fig:earlydiff} we conclude that the profile of the 9.7
$\mu$m silicate features in different diffuse sightlines is very
similar. In Fig.~\ref{fig:StRS_galcent} the difference {between
  our observed silicate features and the Galactic Centre feature
  \citep{Kemper04} is shown}. For both StRS 164 and StRS 354 there are
no strong variations present that exceed the noise. Only StRS 136
shows a small difference with the Galactic {Centre} feature
between about 8 and 9.7 $\mu$m, but this never exceeds 0.15 in the
normalised $\tau$. The two small {absorption} features at about
8.7 and 11.3 $\mu$m in Fig.~\ref{fig:StRS_galcent} (upper panel)
cannot be trusted, because PAHs (Polycyclic Aromatic Hydrocarbons)
have prominent features at those wavelengths and the background
subtraction probably was not accurate enough in this case.

In conclusion, there are only small differences in the shape of the
9.7 $\mu$m silicate feature in the diffuse sightlines and the Galactic
Centre feature, even though these sightlines are separated by up to 77
degrees in {Galactic} longitude (see
Table~\ref{tab:samplesel}).

\subsubsection{Molecular sightlines}
 \label{sec:molsight}
\begin{figure}
\centering
\includegraphics[angle=270,width=8cm]{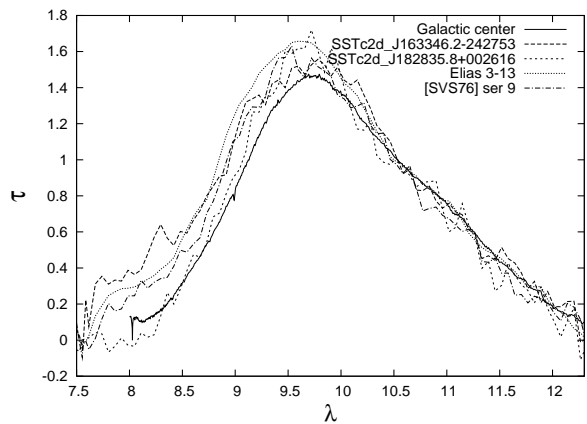}
\caption{The 9.7 $\mu$m silicate features in terms of optical depth
  for the molecular sightlines together with the feature observed
  towards the {Galactic Centre} \citep{Kemper04}. The features
  are normalised to unity at 10.5 $\mu$m.}
\label{fig:earlymol}
\end{figure}

 Figure~\ref{fig:earlymol} shows the extracted features for the
 molecular sightlines in our sample. All features are normalised to
 unity at 10.5 $\mu$m {as was done} for the diffuse sightlines.
 From an eyeball comparison of the 9.7 $\mu$m silicate features in
 Fig.~\ref{fig:earlymol} we conclude that these features in molecular
 sightlines differ substantially from the absorption feature observed
 towards the {Galactic Centre}. The slopes of the long
 wavelength side of the features (i.e. $\lambda >$ 10.5 $\mu$m) are
 the same as that of the Galactic {Centre}. If we assume that
 the profiles are indeed the same from 10.5 $\mu$m onward, then the
 molecular sightlines show excess absorption between 7.5 and 10.5
 $\mu$m. {Alternatively, if we were to normalise the 9.7 $\mu$m
   silicate features at their peak, the difference between them can be
   interpreted as a shift towards shorter wavelengths as compared to
   the diffuse 9.7 $\mu$m silicate features. Also, the long wavelength
   part of the features show less overlap than when they are
   normalised at 10.5 $\mu$m.}

\begin{figure}[!h]
\centering
\includegraphics[angle=270,width=8cm]{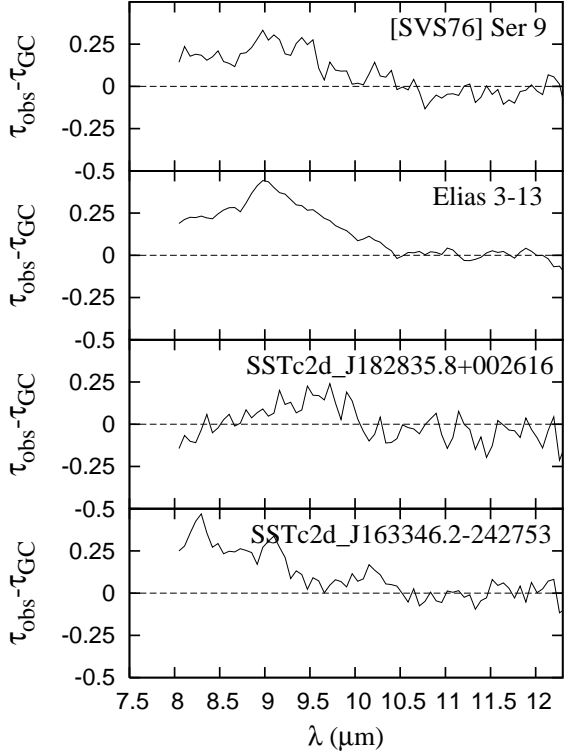}
\caption{The difference between the observed features in the molecular
  sightlines in our sample and the {Galactic Centre} feature
  derived by \citet{Kemper04}. (The {Elias 3-13} spectrum has
  not been corrected for the photospheric gas phase SiO band in this
  figure)}
\label{fig:mol_galcent}
\end{figure}

In Fig.~\ref{fig:mol_galcent} the difference is plotted between the
observed features in the molecular sightlines in our sample and the
{Galactic Centre} feature.  The observed difference between the 9.7
$\mu$m silicate profiles in molecular sightlines and the Galactic
{Centre} sightline can be caused by three effects: (i) The presence of
a photospheric gas phase SiO band in the spectrum of the background
star. (ii) The presence of ices along the line-of-sight. (iii) A
change in the dust properties of the amorphous silicates. We will
discuss these three effects separately below.
\\
~
\\
{(i) Photospheric gas phase SiO band:} Since the spectral type for 3
of the 4 sources is unknown, there might be a small photospheric gas
phase SiO band in the spectra. {SiO is the most {probable}
  source of photospheric spectral structure in cool stars in the 8
  $\mu$m wavelength range}. Indeed, {Elias 3-13} is known to be a K2
giant and, hence, the spectrum of this source will have a weak SiO
band. Careful inspection of the 9.7 $\mu$m features in
Fig.~\ref{fig:10feats2} shows that the photospheric gas phase SiO band
is indeed present in all but one of the molecular sightlines as a blue
shoulder between 7.5 and about 8.5 $\mu$m. Only for
{SSTc2d\_J182835.8+002616 the SiO band seems to be absent}, since
$\tau$ is zero at about 8 $\mu$m, where the SiO band peaks. So for
this source, the observed difference with the Galactic {Centre}
feature {cannot} be explained by the presence of a gas phase SiO band
in the intrinsic spectrum of the background star.

However, Fig.~\ref{fig:mol_galcent} shows that the difference in the
shape of the 9.7 $\mu$m silicate features, in most cases, {cannot} be
explained only by SiO. This is because gas phase SiO normally peaks at
about 8 $\mu$m, whereas Fig.~\ref{fig:mol_galcent} shows features that
peak at about 9--9.5 $\mu$m, except for SSTc2d\_J163346.2$-$242753.

\begin{figure}
\centering
\includegraphics[angle=270,width=7cm]{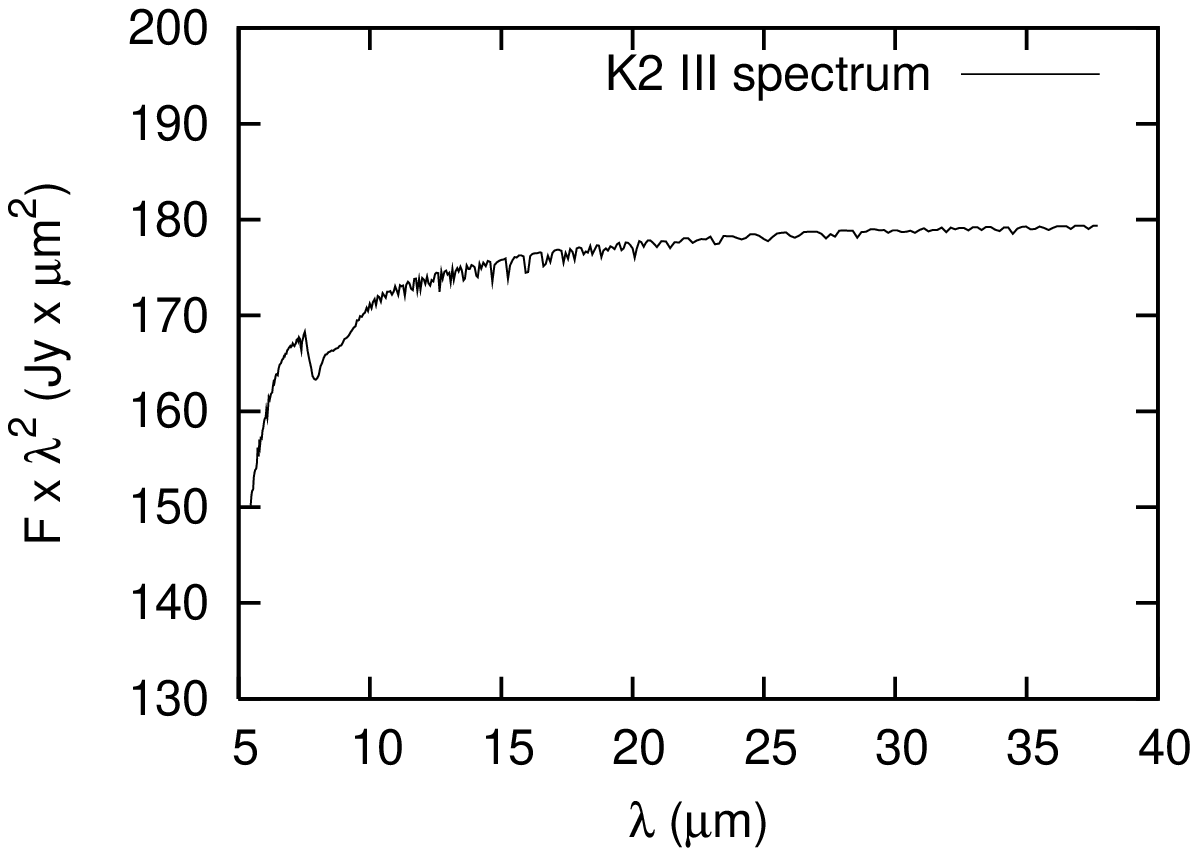}
\includegraphics[angle=270,width=7cm]{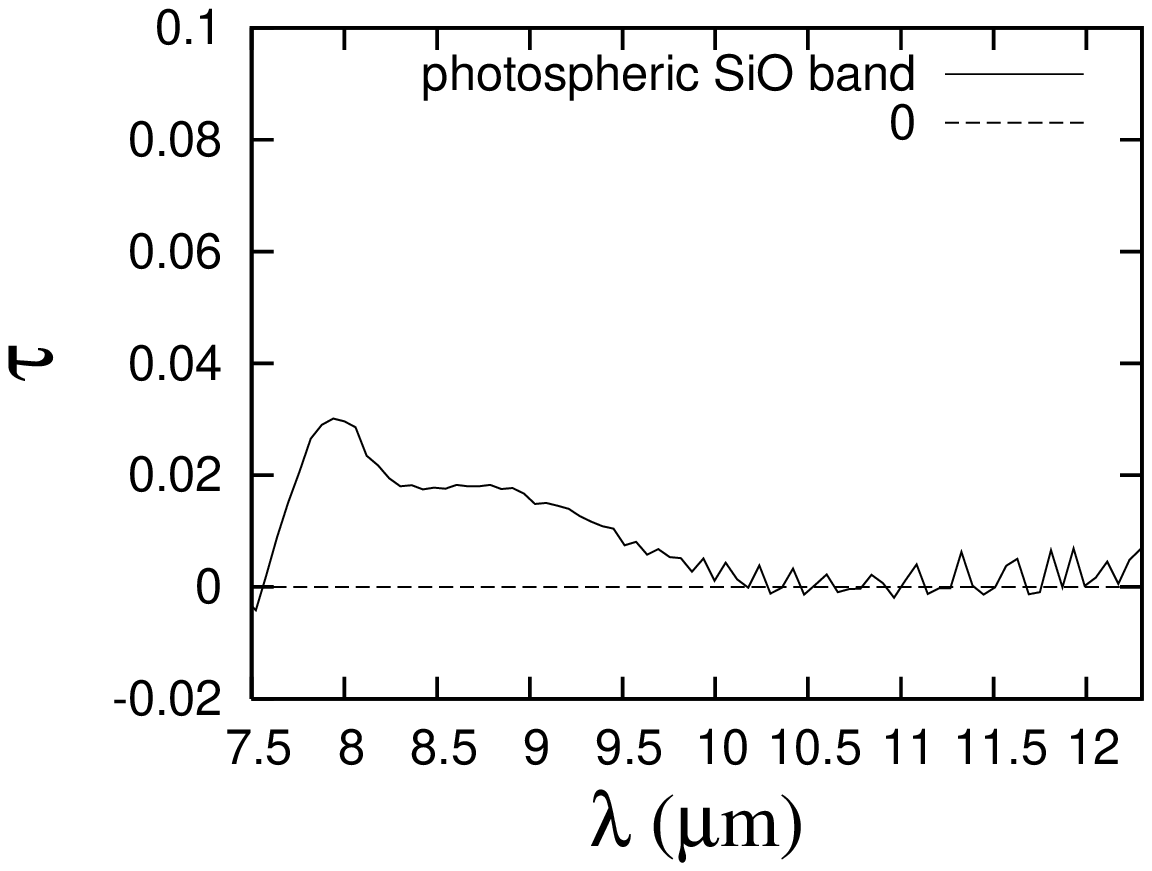}
\caption{\textit{upper}: Model atmosphere spectrum of a K2 giant calculated by L. Decin \citep{Decin97,2004ApJS..154..408D}. \textit{lower}: The corresponding photospheric gas phase SiO band in terms of optical depth. {Longwards of 9 $\mu$m the absorption is dominated by OH.}}
\label{fig:K2III}
\end{figure}

 Since the spectral type of {Elias 3-13} is known we were able
 to extract the photospheric gas phase SiO band for this source using
 a model atmosphere spectrum calculated by {one of us}
 \citep{Decin97,2004ApJS..154..408D} based on MARCS models
 \citep{Gustafsson75,2008A&A...486..951G} (see Fig.~\ref{fig:K2III},
 upper panel).

\begin{figure}
\centering
\includegraphics[angle=270,width=7cm]{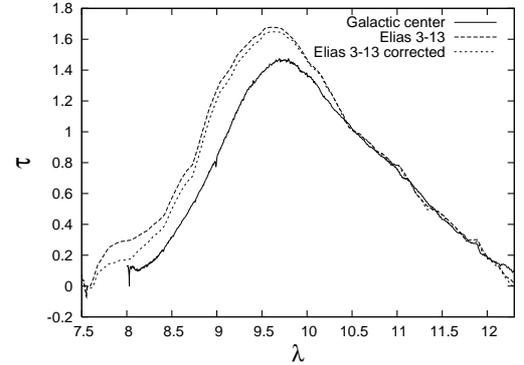}
\caption{The 9.7 $\mu$m silicate feature observed towards
  {Elias 3-13} (dashed line) together with this feature
  corrected for photospheric gas phase SiO (dotted line) using a model
  spectrum calculated by L. Decin
  \citep{Decin97,2004ApJS..154..408D}. The {Galactic Centre}
  feature \citep{Kemper04}) is also plotted (solid line) and is
  representative for diffuse sightlines.}
\label{fig:testElias13}
\end{figure}

The photospheric gas phase SiO band in terms of optical depth was
extracted from this model spectrum by taking a power-law continuum
from 7.4 to 12.3 $\mu$m and using equation 1 (see
Fig.~\ref{fig:K2III}, lower panel). {The resulting optical
  depth profile is mostly due to SiO, but a contribution from OH is
  present at $\lambda \gtrsim 9$ $\mu$m, and OH absorption dominates
  at $\gtrsim 10$ $\mu$m.} Subsequently, we subtracted this optical
depth profile from the observed 9.7 $\mu$m feature. The correction
never exceeds 0.03 in $\tau$. The result is shown in
Fig.~\ref{fig:testElias13}. Correcting for SiO in this spectrum is not
sufficient to explain the difference in the shape of the 9.7 $\mu$m
silicate feature between {Elias 3-13} and the Galactic
{Centre}.
\\
~
\\
{(ii) Ices:} Different types of ices could affect the shape of
the 9.7 $\mu$m silicate feature in molecular sightlines
\citep{Boogert04,Dishoeck04}. {Indeed, \citet{2009ApJ...693L..81M}
  proposed ice growth as cause for the change in extinction law
  between diffuse and molecular sightlines.}
For instance, CH$_{3}$OH ice has a
strong feature at about 9.7 $\mu$m (see also
Fig.~\ref{fig:sst18}). Since the spectra {show no} strong
absorption bands at about 6 $\mu$m and 6.85 $\mu$m, caused by water
ice and CH$_{3}$OH ice together with NH$_{4}^{+}$ ice respectively
(see Fig.~\ref{fig:spectra1}), we {do not} expect the 9.7
$\mu$m silicate features to be affected much by ice features.

For the same reason we {do not} expect a strong H$_{2}$O
libration band at about 12 $\mu$m
\citep{2008ApJ...678..985B}. Moreover, the long wavelength side of the
9.7 $\mu$m silicate features (from about 10.5 $\mu$m onward) in our
molecular sightlines is very similar to that of the diffuse sightlines
indicating that there is indeed no H$_{2}$O libration band present in
any of the spectra.

\begin{figure}
\centering
\includegraphics[angle=270,width=8cm]{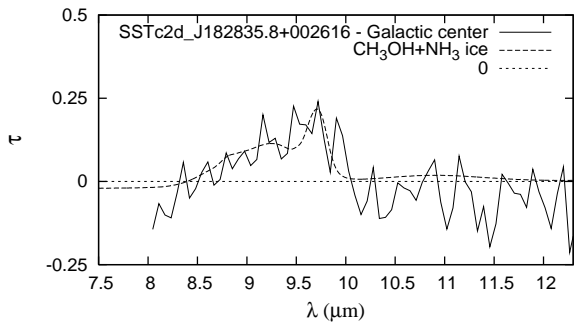}
\caption{The difference between the observed 9.7 $\mu$m feature of
  {SSTc2d\_J182835.8+002616} and the Galactic Centre feature
  derived by \citet{Kemper04} (solid line) together with the
  absorption profile of a mixture of CH$_{3}$OH and CH$_{3}$ ice
  (dashed line).}
\label{fig:SSTc2dJ18..minGal_ice}
\end{figure}

In Fig.~\ref{fig:mol_galcent}, only {SSTc2d\_J182835.8+002616}
shows a feature that peaks at about 9.7 $\mu$m, that may be compatible
with the CH$_{3}$OH ice band. Figure~\ref{fig:SSTc2dJ18..minGal_ice}
shows the difference between the 9.7 $\mu$m feature of
{SSTc2d\_J182835.8+002616} and the Galactic {Centre}
feature together with the absorption profile of a mixture of
CH$_{3}$OH and NH$_{3}$ ice. In this case the observed difference
could be explained by the presence of ices. On the other hand the
spectrum of {SSTc2d\_J182835.8+002616} does not show an
absorption band at 6.85 $\mu$m, caused by CH$_{3}$OH ice together with
NH$_{4}^{+}$ ice.

For {Elias 3-13}, however, the difference between its 9.7
$\mu$m feature and the {Galactic Centre} feature peaks at about
9 $\mu$m even if we correct for the photospheric gas phase SiO
band. Moreover, it is a relatively broad feature covering about 2--3
$\mu$m in wavelength. There are, to the best of our knowledge, no ice
species that produce such an absorption profile.

{(iii) Dust properties of the amorphous silicates:} The shape of the
9.7 $\mu$m silicate absorption feature depends on the dust properties
of the amorphous silicates, such as the size, shape and chemical
composition of the dust grains. The observed increase in absorption
between 7.5 and 10.5 $\mu$m (i.e the short wavelength side of the
feature) could, for instance, be the result of a decrease in the O/Si
ratio in the amorphous silicates, going from an olivine to pyroxene
composition {\citep{OHM_92_cosmicsilicates,2008ApJ...687L..91S}}. {In
  Sect.~\ref{sect:models} we will return to this point.}

{From the discussion above we conclude that it is not likely
  that gas-phase SiO absorption or ice absorption are the dominant
  cause of the difference in the band shape of the 9.7 $\mu$m silicate
  absorption band in diffuse and molecular lines-of-sight. Changes in
  the dust composition may be important in explaining the difference.}

\subsection{Comparison of the 18 $\mu$m silicate profiles}

\begin{figure}[!h]
\centering
\includegraphics[angle=270,width=8cm]{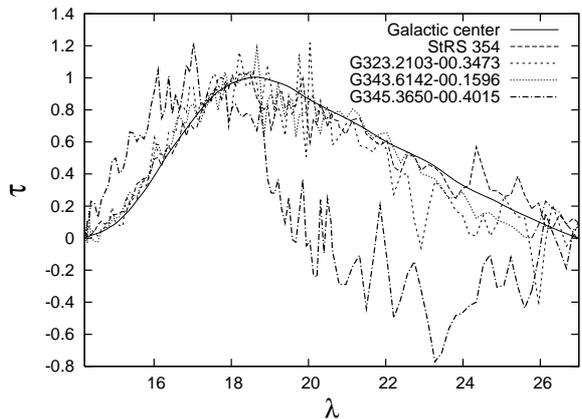}
\caption{The 18 $\mu$m silicate absorption features observed in 4
  diffuse sightlines together with {the modelled profile based on the 
  9.7 $\mu$m fit of the Galactic Centre sightline} \citep{Min07}. The features are normalised at
  their peak. {Noise levels vary with wavelength and are
    typically 0.03 in $\tau$ at the shortest wavelengths, and increase
    to 0.10 in $\tau$ for the longest wavelength. {The anomalous feature seen towards G345.3650$-$00.4015 is plotted with a thicker line than the features in the other sightlines}.}}
\label{fig:18micfeats}
\end{figure}

The 18 $\mu$m silicate absorption profiles {are} extracted for
4 diffuse sightlines (see
Sect.~\ref{sect:extraction}). {Because we could analyse the
9.7 $\mu$m feature in only one of these sightlines, it is not possible
for us to use the feature strength ratio between the two features
as a tool to put constraints on grain properties, such as composition 
and size \citep[e.g.~][]{OHM_92_cosmicsilicates}.} 
Figure~\ref{fig:18micfeats} shows {the 18 $\mu$m}
profiles together with {a modelled Galactic Centre 18 $\mu$m profile, 
based on the grain properties derived from the 9.7 $\mu$m fit in the same
sightline} \citep{Min07}. The features are normalised to unity at
their peak.

{The 18 $\mu$m silicate profile of G345.3650$-$00.4015 is the only one
  that does not agree with the extrapolated Galactic Centre fit within
  the noise and we have found no reasonable explanation for this
  particular profile being so different. {For this sightline,
    we considered a range of olivine and pyroxene silicates with
    different Mg/Fe ratios, as well as Na- and Al-containing
    silicates. For all these species we considered spherical and
    irregularly-shaped particles, none of which reproduce the required
    blueness of this anomalous feature.  Although we suspect that a
    compositional effect causes the shift in wavelength for this
    feature, the carrier remains unidentified.}  However, the
  {18 $\mu$m} features {in the other sightlines} are
  very similar to that of the extrapolated Galactic Centre fit and
  \citet{Min07} show that this profile is also consistent with the 18
  $\mu$m silicate absorption feature observed towards WR 98a.} So,
except for one source, the 18 $\mu$m silicate absorption features in
different diffuse sightlines are very similar and the extrapolated
{Galactic Centre} fit seems representative for the diffuse ISM.

\subsection{{Summary of observed trends}}

For lines-of-sight passing through the diffuse ISM, there is a tight
linear correlation between the strength of the 9.7 $\mu$m silicate
absorption feature, $\tau_{9.7}$, and the {near-infrared} colour excess,
$E$(J$-$K). Moreover, the shape of the observed 9.7 $\mu$m silicate
features in different diffuse sightlines and the Galactic
{Centre} feature are strikingly similar (see
Fig.~\ref{fig:earlydiff}). Only StRS 136 shows a small difference with
the Galactic Centre feature between about 8 and 9.7 $\mu$m of up to
0.13 in $\tau$. It could be that this line-of-sight not only probes
diffuse material but also a small amount of molecular material. If
this is indeed the case, then the Galactic {Centre} feature is
representative for the diffuse ISM even though this feature was
extracted using a very different method \citep{Kemper04} than applied
here. This is remarkable, since the Galactic {Centre}
line-of-sight not only passes through diffuse material, but also some
molecular cloud material \citep{R_88_GC,Whittet03}.

Our analysis of the 18 $\mu$m silicate absorption feature shows the
same similarity between diffuse sightlines and the {Galactic
  Centre}. Except for one {source} (i.e. MSXLOS17) the 18
$\mu$m profiles are consistent with the extrapolated Galactic
{Centre} fit from \citet{Min07}.

We conclude that the diffuse ISM silicates probed by the diffuse
lines-of-sight studied in this paper and those sampled towards the
{Galactic Centre} are remarkably similar. Only the
$\tau_{9.7}$/$E$(J$-$K) of the {Galactic Centre} sightline
deviates from the general behaviour.

{A different picture appears when looking at lines-of-sight
  passing through molecular cloud material. The $\tau_{9.7}$ versus
  $E$(J$-$K) relationship breaks down drastically and we observe small
  variations in the shape of the 9.7 $\mu$m silicate absorption
  features. These small changes may be attributed to differences in
  dust properties, but other effects may also play a role (see above).
  Nevertheless, the observations show that the dust properties in
  molecular clouds must be different from those in the diffuse ISM.}

{To determine whether} the change in the $\tau_{9.7}$ versus
$E$(J$-$K) relationship is correlated to the change in the shape of
the 9.7 $\mu$m silicate feature, we have measured the 9.7 $\mu$m
silicate band shape of three sources in Fig.~\ref{fig:tauvsEJK} with a
wide range in {$\tau_{9.7}$/$E$(J$-$K)}.{ However, we
  found that better quality data than available to us in this study
  are needed to come to firm conclusions about such a possible
  relation.}

In the following we will consider possible explanations for the
observed variations in both the $\tau_{9.7}$ versus $E$(J$-$K) diagram
and the shape of the 9.7 $\mu$m silicate profile. We model the
$\tau_{9.7}$ versus $E$(J$-$K) relationship assuming a range of dust
properties; then we confront these models with the observations.

\section{{Model calculations}}
\label{sect:models}

To investigate how the $\tau_{9.7}$ versus $E$(J$-$K) relationship and
the 9.7 $\mu$m silicate absorption profile depend on different dust
properties we start with the best fit found by \citet{Min07} for the
{Galactic Centre} feature. This fit uses an MRN
\citep{Mathis77} size distribution (i.e. $n(a) \propto a^{-3.5}$),
between grain sizes of 0.005 and 0.25 $\mu$m. The shape of the
particles is represented by a distribution of hollow spheres (DHS),
where $f_{\mathrm{max}}$ represents the size of the cavities in the grains and
is a measure for the irregularity of the particles, {and a larger
$f_{\mathrm{max}}$ indicates that the grains are more porous and/or irregular}
\citep{2005A&A...432..909M}. A DHS model with $f_{\mathrm{max}}=1$
simulates highly irregular particles, while a DHS model with
$f_{\mathrm{max}}=0.1$ simulates almost spherical particles. The best
fit model from \citet{Min07} uses $f_{\mathrm{max}}=0.7$. The chemical
composition of this best fit model is given in
Table~\ref{tab:comp_galcent}.

\begin{table}
\centering
\begin{tabular}{lll}
\hline \\
Silicates & Molecule & Mass fraction (\%)\\
\hline\\
Amorphous olivine & MgFeSiO$_{4}$ & 13.8 \\
 & Mg$_{2}$SiO$_{4}$ & 38.3 \\
Amorphous pyroxene & MgSiO$_{3}$ & 42.9 \\
Amorphous Na/Al pyroxene & NaAlSi$_{2}$O$_{6}$ & 1.8 \\
Crystalline olivine (Forsterite) & Mg$_{2}$SiO$_{4}$ & 0.6\\
\hline\\
Other & & \\
\hline\\
Crystalline silicon carbide & SiC & 2.6 \\
\hline\\
\end{tabular}
\caption{The chemical composition in terms of mass fraction in percent for the best fit model of the {Galactic Centre} feature by \citet{Min07}. This model uses a distribution of hollow spheres (DHS) with f$_{max}=0.7$ and an MRN size distribution with index $-$3.5 between a grain size of 0.005 and 0.25 $\mu$m.}
\label{tab:comp_galcent}
\end{table}

By changing the parameters in this model we are able to investigate
the effects of grain size, grain shape and chemical composition of
different types of silicates and silicon carbide on the 9.7 $\mu$m
silicate absorption profiles.

To model the relationship between $\tau_{9.7}$ and $E$(J$-$K) we have
included a fraction of graphite (C) in such a way that for the
{Galactic Centre} the diffuse ISM correlation \citep{Chiar07}
is obtained. Note that this correlation is not valid for the
{Galactic Centre} sightline in particular (see
Fig.~\ref{fig:tauvsEJK}), but the 9.7 $\mu$m silicate absorption
profile is representative for the diffuse ISM and thus we will
consider it as a typical diffuse sightline. A mass
fraction of about 31 \% of graphite was needed to meet this
requirement. Assuming a solar abundance of carbon, this requires about
65 \% of the carbon atoms to be in the form of graphite, which is a
relatively high percentage. {This number should be treated as 
an upper limit; when metallic iron or \emph{dirty silicates} are present 
in the diffuse ISM, they will contribute to the NIR opacity, and thus
decrease the contribution from graphite.}

To verify if this model also reproduces the standard extinction law at
optical wavelengths the corresponding value for
R$_{V}$=$A_{\mathrm{V}}$/$E$(B$-$V) was calculated. The result was
3.28, which is close to the value for the diffuse ISM of about 3.1
\citep{1975A&A....43..133S,1978ApJ...223..168S}.  We consider the
following dust properties and investigate their effect on the
$\tau_{9.7}$ versus $E$(J$-$K) relationship: (1) The size distribution
of the dust grains, (2) The shape distribution of the dust grains, (3)
The chemical composition of the dust grains. In the following we will
consider graphites and silicates as separate grains and we will assume
that they have the same size and shape distribution.

\subsection{Grain size}
\begin{figure}
\centering
\includegraphics[angle=270,width=6cm]{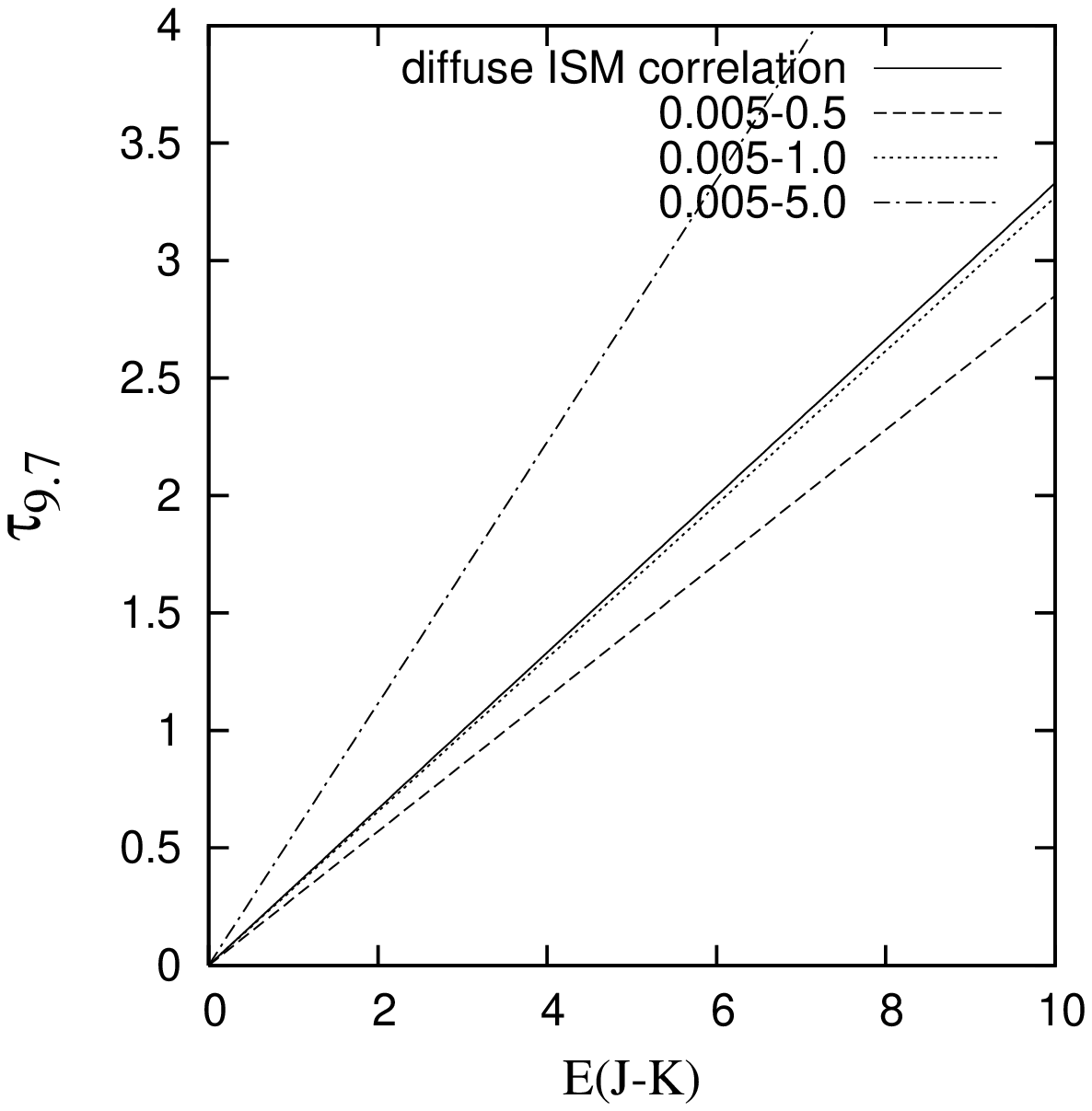}
\includegraphics[angle=270,width=9cm]{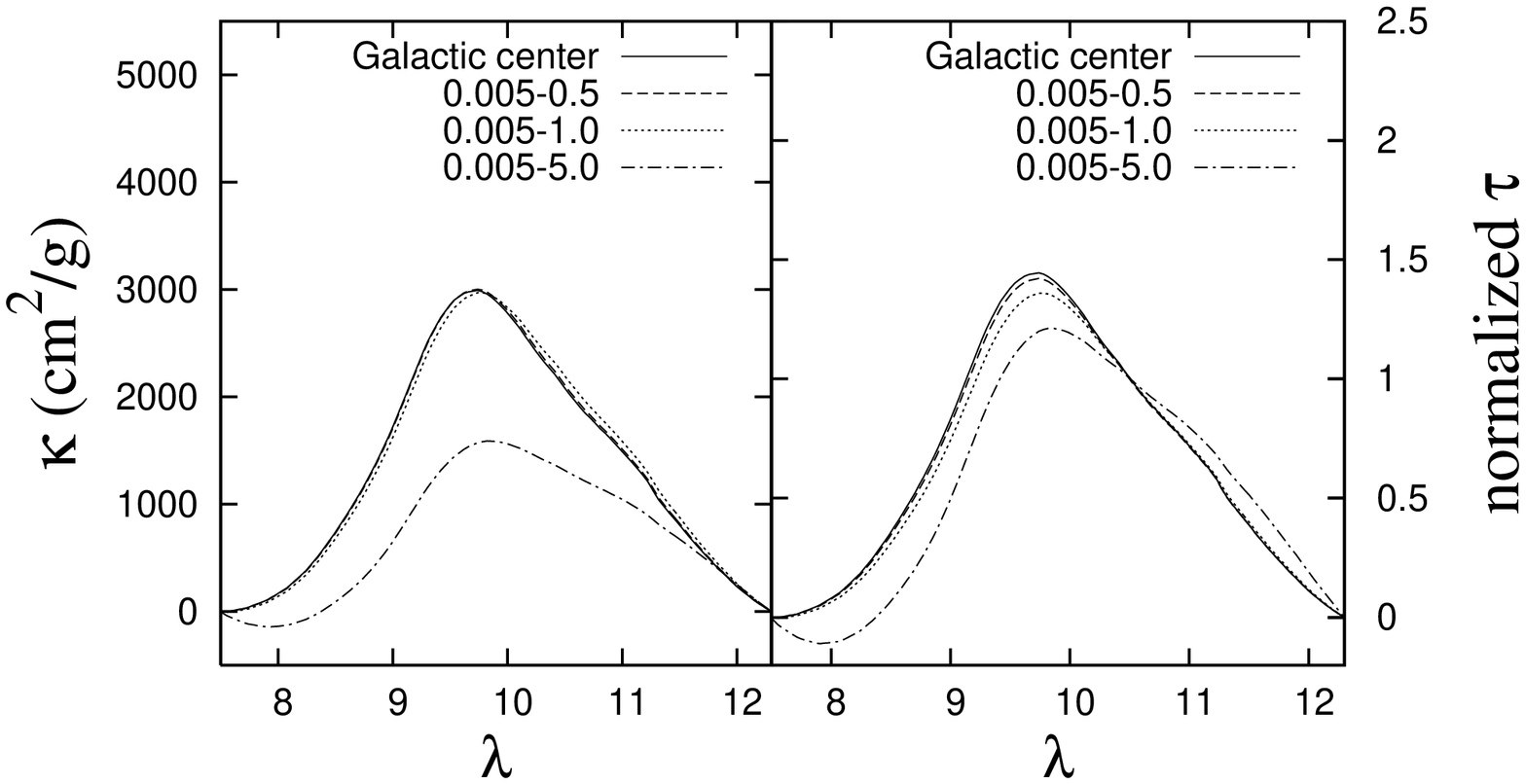}
\caption{\textit{upper:} The effect of increasing the maximum grain
  size on the the $\tau_{9.7}$ versus $E$(J$-$K) relationship. The
  minimum and maximum grains size are given in
  $\mu$ms. \textit{lower:} The effect of increasing the maximum grain
  size on the shape of the 9.7 $\mu$m silicate feature. The left panel
  shows the calculated 9.7 $\mu$m silicate absorption profiles in
  terms of the mass extinction coefficient and the right panel shows
  the same profiles normalised to 1 at 10.5 $\mu$m {as was
    done} with the observations. The solid line in both panels
  represent the best fit for the {Galactic Centre} feature from
  \citet{Min07}.}
\label{fig:size}
\end{figure}

{Due to the high density, dust grains are likely to grow in
  molecular clouds. In order to model this, we increase} the maximum
grain size $r_{\mathrm{max}}$ while keeping $r_{\mathrm{min}}$
constant, and calculate $\tau_{9.7}$/$E$(J$-$K) as well as the shape
of the 9.7 $\mu$m silicate feature. The result is shown in
Fig.~\ref{fig:size}.  At first, when the maximum grain size is
increased, $\tau_{9.7}$/$E$(J$-$K) decreases down to about 0.28, but,
as $r_{\mathrm{max}}$ {exceeds} $\sim$0.6 $\mu$m, it starts
increasing. This behaviour is dominated by the change in $E$(J$-$K) due
to growth of the graphite grains. With this model, however, we are not
able to explain the observed flattening of the $\tau_{9.7}$ versus
$E$(J$-$K) relationship, because our model curves do not reach the
lower right part of the $\tau_{9.7}$, $E$(J$-$K) diagram, where the
most reddened molecular lines-of-sight are located (see
Fig.~\ref{fig:tauvsEJK}).

The shape of the 9.7 $\mu$m silicate feature does not change much as
the maximum grain size is increased up to 1.0 $\mu$m. When
$r_{\mathrm{max}}$ {exceeds this value} the 9.7 $\mu$m silicate
profile weakens and shifts towards longer wavelengths. This is in
contrast with our observations which show slight excess absorption on
the short wavelength side of the feature (see
Fig.~\ref{fig:earlymol}). So, if grain growth is indeed taking place,
it is limited to a grain size of about 1.0 $\mu$m, {and cannot
explain the observed variations.}

\subsection{Grain shape}
\begin{figure}
\centering
\includegraphics[angle=270,width=6cm]{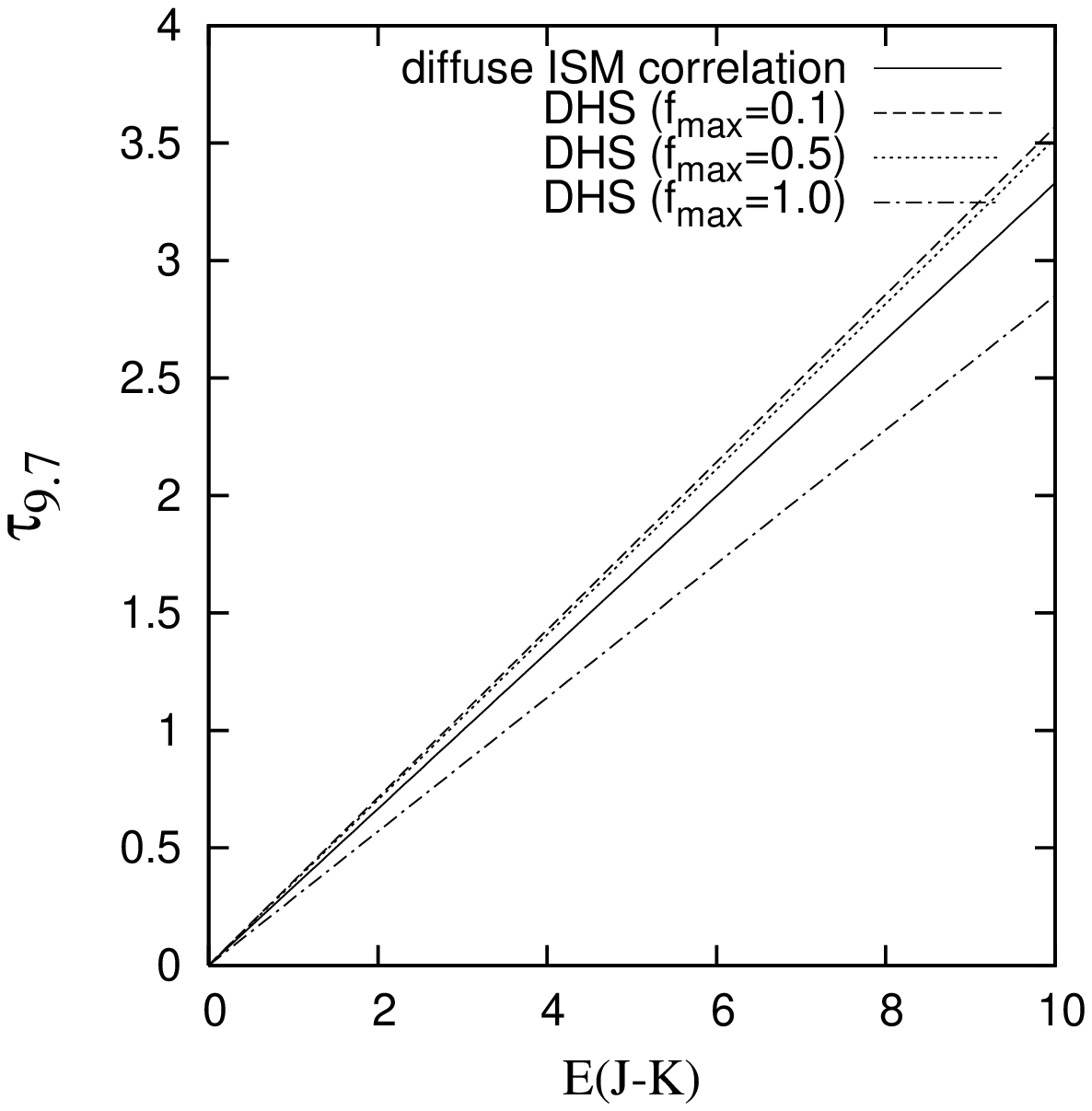}
\includegraphics[angle=270,width=9cm]{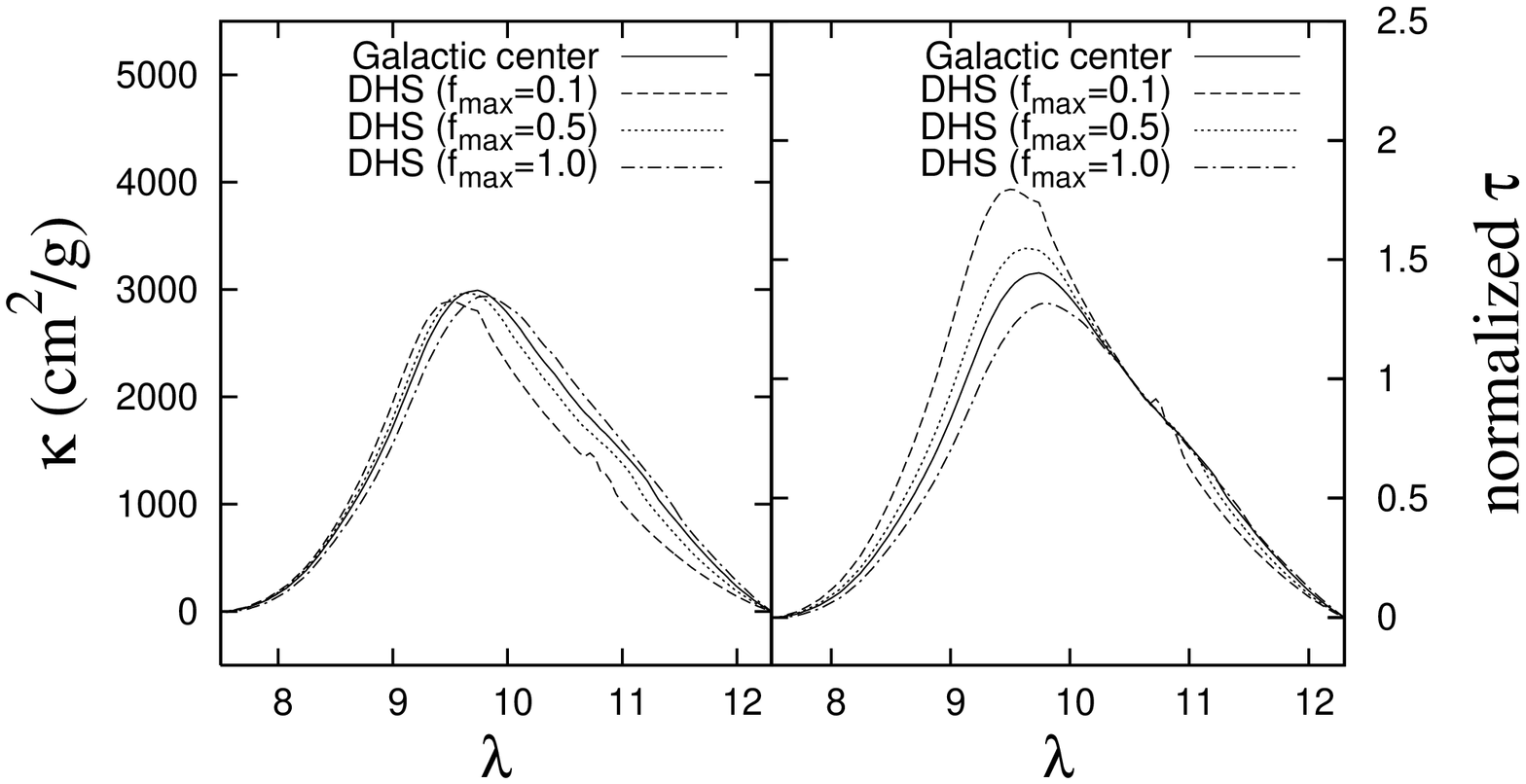}
\caption{\textit{upper:} The effect of changing the grain shape on the
  the $\tau_{9.7}$ versus $E$(J$-$K) relationship. The shape of the
  particles is represented by a distribution of hollow spheres (DHS),
  where f$_{max}$ represents the size of the cavities in the grains
  and is a measure for the irregularity of the particles (f$_{max}$=1
  means highly irregular particles, f$_{max}$=0.1 means almost
  spherical particles). The Galactic {Centre} fit by
  \citet{Min07} uses f$_{max}$=0.7. \textit{lower:} The effect
  changing the maximum grain shape on the 9.7 $\mu$m silicate
  profile. The left panel shows the calculated 9.7 $\mu$m silicate
  absorption profiles in terms of the mass extinction coefficient and
  the right panel shows the same profiles normalised to 1 at 10.5
  $\mu$m {as was done} with the observations. The solid line in
  both panels represent the best fit for the {Galactic Centre}
  feature from \citet{Min07}.}
\label{fig:shape}
\end{figure}

Figure~\ref{fig:shape} (upper panel) shows how the $\tau_{9.7}$ versus
$E$(J$-$K) relationship changes as a function of f$_{max}$. {With 
increased sphericity, the slope of the relation decreases towards the 
value observed in molecular sightlines (see
Fig.~\ref{fig:tauvsEJK}).}
The effect is small, {however,} and certainly not strong enough to explain the observed
break down of this relationship in the observations.

The lower panel of Fig.~\ref{fig:shape} shows the effect of changing
the grain shape on the 9.7 $\mu$m silicate profile. Comparing this to
the observations (Fig.~\ref{fig:earlymol}), the observed variations in
molecular sightlines might be explained if the dust grains in
molecular clouds are more regular than in the diffuse ISM.

\subsection{Chemical composition}

\begin{figure}
\centering
\includegraphics[angle=270,width=6cm]{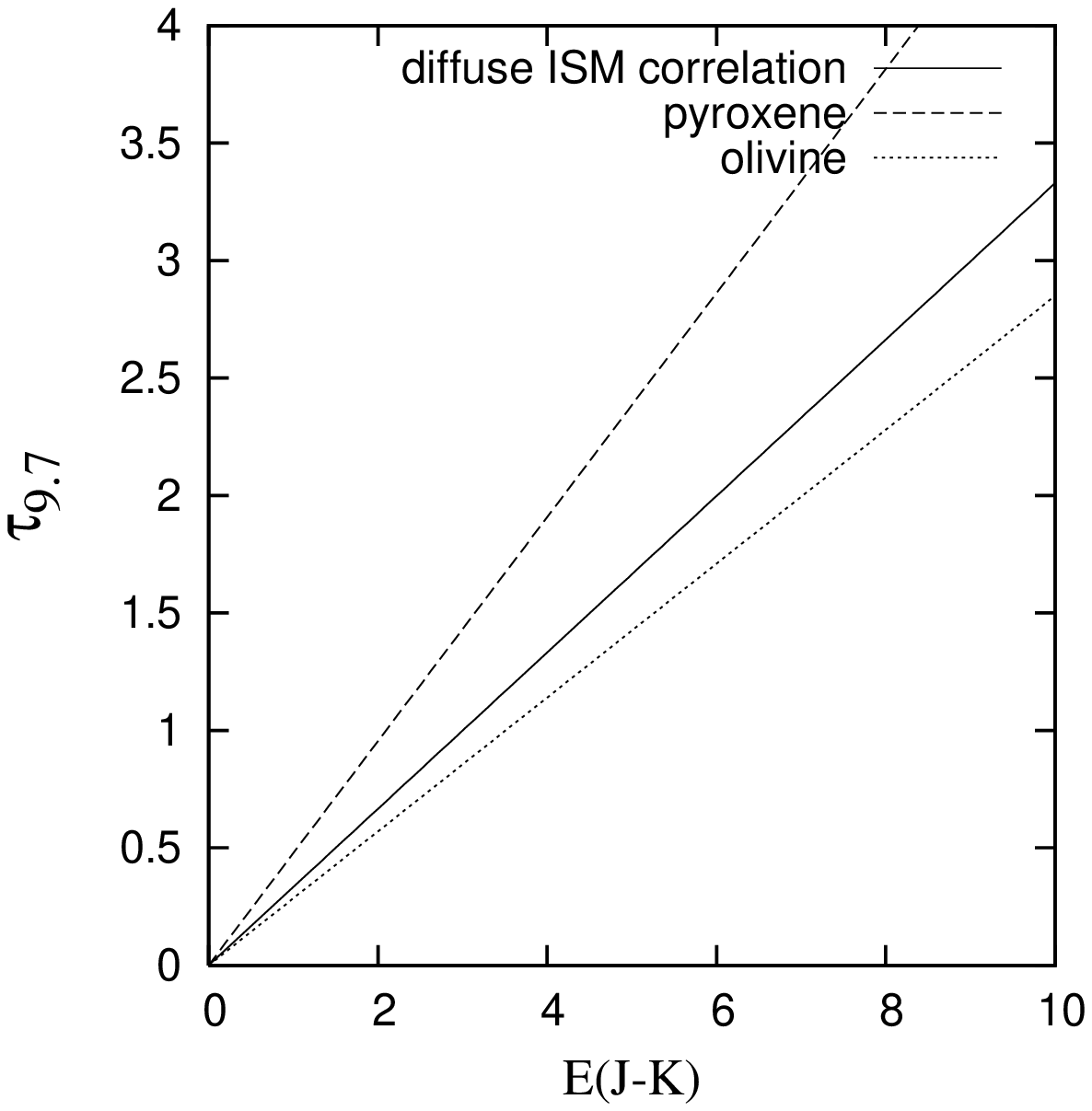}
\includegraphics[angle=270,width=9cm]{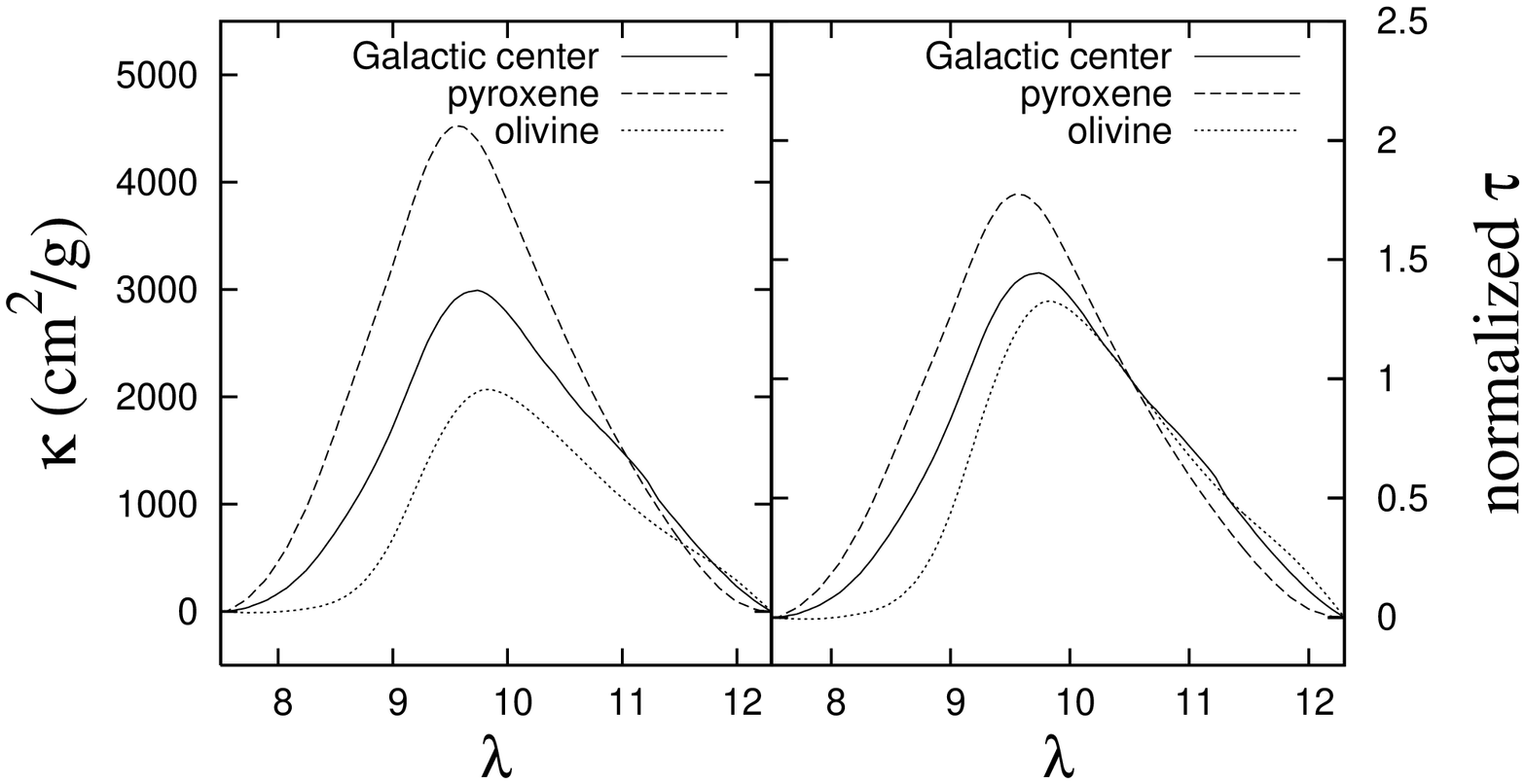}
\caption{\textit{upper:} The $\tau_{9.7}$ versus $E$(J$-$K)
  relationship for the chemical composition of the best fit
  \citet{Min07} found for the {Galactic Centre} together with a
  mass fraction of 31 \% of graphite (solid line). The dashed and
  dotted lines represent a model with the same amount of graphite
  together with a pure magnesium rich amorphous pyroxene
  (i.e. MgSiO$_{3}$) and magnesium rich amorphous olivine
  (i.e. Mg$_{2}$SiO$_{4}$), respectively. \textit{lower:} The 9.7
  $\mu$m silicate absorption profiles for the above models. The left
  panel shows the calculated 9.7 $\mu$m silicate absorption profiles
  in terms of the mass extinction coefficient and the right panel
  shows the same profiles normalised to 1 at 10.5 $\mu$m {as
    was done} with the observations. The solid line in both panels
  represent the best fit for the {Galactic Centre} feature from
  \citet{Min07}.}
\label{fig:comp}
\end{figure}

Interstellar silicate dust is mostly a mixture of amorphous olivines
and pyroxenes
\citep[e.g.][]{1974ApJ...192L..15D,1975A&A....45...77G,Kemper04,Min07,Chiar07}.
Figure~\ref{fig:comp} shows the effect of changing the chemical
composition of the silicates on the $\tau_{9.7}$ versus $E$(J$-$K)
relationship. This relationship is plotted for the chemical
composition of the best fit \citet{Min07} found for the {Galactic
  Centre}, {where we used graphite instead of amorphous carbon,
  at a mass fraction of 31 \%} {\citep[solid
  line;][]{2003ApJ...598.1026D}}. The dashed and dotted lines
represent {models} with the same amount of graphite, together with a
type of amorphous pyroxene
{\citep[i.e.~MgSiO$_3$;][]{DBH_95_glasses}} {only; or graphite
  and a type of} amorphous olivine
{\citep[i.e.~Mg$_2$SiO$_4$;][]{HS_96_opacities}} {only,}
respectively. Going from a pyroxene to olivine composition
$\tau_{9.7}$/$E$(J$-$K) decreases, but even assuming a pure olivinic
composition is not sufficient to explain the observations.
{The best fit to the diffuse ISM feature towards the Galactic
  Centre contains Mg-rich end-members of silicates only \citep{Min07},
  and the contribution to the NIR opacity of these silicates is
  negligible. The presence of iron-containing silicates, or
  \emph{dirty silicates}, would increase the contribution to the NIR
  opacity due to silicates, and thus decrease the amount of graphite
  required.}

Figure~\ref{fig:comp} (lower panel) shows the effect of changing the
silicate composition on the shape of the 9.7 $\mu$m silicate
feature. The change in shape of the 9.7 $\mu$m silicate feature that
we observe could be explained if the silicate grains in molecular
clouds contain relatively more amorphous pyroxene.  This would
increase the $\tau_{9.7}$ versus $E$(J$-$K) relationship slightly, in
contrast to the observations, but, since the observed variations in
the 9.7 $\mu$m silicate profile are small, the effect on the
$\tau_{9.7}$ versus $E$(J$-$K) relationship is probably also small.

\section{{Discussion}}

\subsection{Silicates ($\tau_{9.7}$) as cause of the variations}
The observed breakdown of the $\tau_{9.7}$ versus $E$(J$-$K)
relationship in molecular clouds could either be caused by a decrease
of $\tau_{9.7}$ or an increase of $E$(J$-$K). In order to explain the
behaviour of the molecular cloud sightlines in terms of variations in
$\tau_{9.7}$, the silicate optical depth must decrease by at least a
factor of 2 compared to the diffuse ISM silicates. Our model
calculations show that such a large change in $\tau_{9.7}$ is hard to
achieve without substantial changes in the shape of the 9.7 $\mu$m
silicate absorption band. The observed modest band shape variations in
molecular cloud sightlines suggest that variations in $\tau_{9.7}$ can
not be the dominant cause for the difference between diffuse and
molecular sightlines. Instead, variations in $E$(J$-$K) may be
responsible.

The observed small changes in the silicate band shape are probably not
caused by grain growth, unless grain growth causes more spherical
grain shapes (see Sect.~\ref{sect:models}). {Note} that grain
growth can still take place, but it is limited to a grain size of
about 1 $\mu$m, since above this grain size the 9.7 $\mu$m silicate
profile changes in a way that is inconsistent with the observations.

In the above models, we have not considered chemically inhomogeneous
grains. For instance, metal inclusions can have an effect on the
silicate absorption.  Such changes should
  specifically affect the silicate optical depth, as the NIR
  extinction law remains unchanged over a large range of column
  densities \citep{2007ApJ...664..357R}.

\subsection{$E$(J$-$K) as cause of the variations}

Since we {did not find a} model that can explain the breakdown
of the $\tau_{9.7}$ versus $E$(J$-$K) relationship in terms of
variations in $\tau_{9.7}$, we consider it more likely that this is
due to changes in $E$(J$-$K). There are several ways to achieve such
variations without affecting the shape and strength of the 9.7 $\mu$m
silicate absorption band.

The effect of grain growth and shape changes in graphite grains on the
$\tau_{9.7}$ versus $E$(J$-$K) relationship has been investigated in
Sect.~\ref{sect:models}. Grain growth can be the answer \emph{if}
silicates do not grow beyond about 1 $\mu$m and at the same time this
growth should increase the $E$(J$-$K) by up to a factor of 4. It is
unclear what could be the cause for such a large increase in
$E$(J$-$K) in combination with modest silicate grain growth. Possibly,
inhomogeneous aggregates could cause such behaviour.

In our model, the graphites and the silicates are considered to be
separate grains, but in reality they might form composite
grains. Moreover, our model does not include possible metallic iron
grains, which could cause extinction in the {near-infrared}.

{The $E$(J$-$K) in molecular clouds could also increase by the
  formation of a dust species that is less abundant in the diffuse ISM
  and has a high opacity at near-infrared wavelengths, e.g.~FeS, which
is thought to form in molecular clouds \citep{2003MNRAS.341..657S}.}

\section{Conclusions}

In order to study the dust properties in different environments in the
ISM, spectra were selected from the Spitzer archive of highly
{reddened stars} in both diffuse and molecular sightlines. We
have analysed the 9.7 and 18 $\mu$m silicate absorption features in
these spectra separately. Our main results are:
\begin{itemize}
\item{For diffuse lines-of-sight the strength of the 9.7 $\mu$m
    silicate feature represented by the optical depth at about 9.7
    $\mu$m ($\tau_{9.7}$) shows a tight linear correlation with the
    near infrared colour excess{, $E$(J$-$K)}. However, this
    correlation breaks down for molecular sightlines
    \citep{Chiar07}. The measurements for the spectra analysed in
    {this paper} are consistent with this finding.}
\item{The shape of the 9.7 $\mu$m silicate feature is {very}
    similar for the three diffuse sightlines investigated in this
    study. Their shape is, within errors, also similar to that
    observed in the line-of-sight towards the {Galactic Centre}
    \citep{Kemper04}. We conclude that we only observe small
    variations {less than 0.15 in normalised $\tau$} in the 9.7
    $\mu$m silicate band shape in diffuse lines-of-sight, and that the
    {Galactic Centre} line-of-sight is representative for the
    diffuse ISM in general. {The shape of the 18 $\mu$m
      silicate absorption feature is also, {within errors},
      found to be invariable for the diffuse sightlines.} The
    situation is different for lines-of-sight passing through
    molecular clouds: we observe small variations on the short
    wavelength side of the 9.7 $\mu$m silicate band.}
\item{Since both the shape of the 9.7 $\mu$m silicate feature and the
    relationship between $\tau_{9.7}$ and $E$(J$-$K) change for molecular
    sightlines {as} compared to diffuse sightlines, we
    conclude that the dust properties in molecular clouds are
    different from the diffuse ISM.}

\item{{Based} on comparison with theoretically calculated profiles, we
    have ruled out grain growth beyond about 1 $\mu$m as being
    responsible for the changes in the 9.7 $\mu$m profile. The
    observations could be explained if the dust grains are more
    spherical in molecular clouds than in the diffuse ISM, but the
    underlying process is not understood. }
\item{We propose that the flattening of the $\tau_{9.7}$ versus $E$(J$-$K)
    relationship is caused by an increase in $E$(J$-$K) in molecular
    clouds. The mechanism that causes this, however, is not yet
    clear. Further research is necessary to investigate whether it can
    be caused by grain growth.}
\end{itemize}


\begin{thebibliography}{52}
\expandafter\ifx\csname natexlab\endcsname\relax\def\natexlab#1{#1}\fi

\bibitem[{{Boogert} \& {Ehrenfreund}(2004)}]{Boogert04}
{Boogert}, A.~C.~A. \& {Ehrenfreund}, P. 2004, in Astronomical Society of the
  Pacific Conference Series, Vol. 309, Astrophysics of Dust, ed. A.~N. {Witt},
  G.~C. {Clayton}, \& B.~T. {Draine}, 547

\bibitem[{{Boogert} {et~al.}(2008){Boogert}, {Pontoppidan}, {Knez}, {Lahuis},
  {Kessler-Silacci}, {van Dishoeck}, {Blake}, {Augereau}, {Bisschop},
  {Bottinelli}, {Brooke}, {Brown}, {Crapsi}, {Evans}, {Fraser}, {Geers},
  {Huard}, {J{\o}rgensen}, {{\"O}berg}, {Allen}, {Harvey}, {Koerner}, {Mundy},
  {Padgett}, {Sargent}, \& {Stapelfeldt}}]{2008ApJ...678..985B}
{Boogert}, A.~C.~A., {Pontoppidan}, K.~M., {Knez}, C., {et~al.} 2008, \apj,
  678, 985

\bibitem[{Bouwman {et~al.}(2001)Bouwman, Meeus, {de Koter}, Hony, Dominik, \&
  Waters}]{BMD_01_processing}
Bouwman, J., Meeus, G., {de Koter}, A., {et~al.} 2001, \aap, 375, 950

\bibitem[{{Bowey} {et~al.}(1998){Bowey}, {Adamson}, \&
  {Whittet}}]{1998MNRAS.298..131B}
{Bowey}, J.~E., {Adamson}, A.~J., \& {Whittet}, D.~C.~B. 1998, \mnras, 298, 131

\bibitem[{{Chiar} {et~al.}(2007){Chiar}, {Ennico}, {Pendleton}, {Boogert},
  {Greene}, {Knez}, {Lada}, {Roellig}, {Tielens}, {Werner}, \&
  {Whittet}}]{Chiar07}
{Chiar}, J.~E., {Ennico}, K., {Pendleton}, Y.~J., {et~al.} 2007, \apjl, 666,
  L73

\bibitem[{{Chiar} \& {Tielens}(2006)}]{Chiar06}
{Chiar}, J.~E. \& {Tielens}, A.~G.~G.~M. 2006, \apj, 637, 774

\bibitem[{Compi\`egne {et~al.}(2010)Compi\`egne, Verstraete, Jones, Bernard,
  Boulanger, Flagey, {Le Bourlot}, Paradis, \& Ysard}]{CVJ_10_dustSED}
Compi\`egne, M., Verstraete, L., Jones, A., {et~al.} 2010, \aap, in press

\bibitem[{{Day}(1974)}]{1974ApJ...192L..15D}
{Day}, K.~L. 1974, \apjl, 192, L15

\bibitem[{{Decin} {et~al.}(1997){Decin}, {Cohen}, {Eriksson}, {Gustafsson},
  {Huygen}, {Morris}, {Plez}, {Sauval}, {Vandenbussche}, \&
  {Waelkens}}]{Decin97}
{Decin}, L., {Cohen}, M., {Eriksson}, K., {et~al.} 1997, in ESA Special
  Publication, Vol. 419, The first ISO workshop on Analytical Spectroscopy, ed.
  A.~M. {Heras}, K.~{Leech}, N.~R. {Trams}, \& M.~{Perry}, 185

\bibitem[{{Decin} {et~al.}(2004){Decin}, {Morris}, {Appleton}, {Charmandaris},
  {Armus}, \& {Houck}}]{2004ApJS..154..408D}
{Decin}, L., {Morris}, P.~W., {Appleton}, P.~N., {et~al.} 2004, \apjs, 154, 408

\bibitem[{{Demyk} {et~al.}(2001){Demyk}, {Carrez}, {Leroux}, {Cordier},
  {Jones}, {Borg}, {Quirico}, {Raynal}, \& {d'Hendecourt}}]{Demyk01}
{Demyk}, K., {Carrez}, P., {Leroux}, H., {et~al.} 2001, \aap, 368, L38

\bibitem[{{Demyk} {et~al.}(2000){Demyk}, {Dartois}, {Wiesemeyer}, {Jones},
  {D'Hendecourt}, {Jourdain de Muizon}, \& {Heras}}]{Demyk00}
{Demyk}, K., {Dartois}, E., {Wiesemeyer}, H., {et~al.} 2000, in ESA Special
  Publication, Vol. 456, ISO Beyond the Peaks: The 2nd ISO Workshop on
  Analytical Spectroscopy, ed. A.~{Salama}, M.~F. {Kessler}, K.~{Leech}, \&
  B.~{Schulz}, 183

\bibitem[{{Demyk} {et~al.}(1999){Demyk}, {Jones}, {Dartois}, {Cox}, \&
  {D'Hendecourt}}]{Demyk99}
{Demyk}, K., {Jones}, A.~P., {Dartois}, E., {Cox}, P., \& {D'Hendecourt}, L.
  1999, \aap, 349, 267

\bibitem[{Dorschner {et~al.}(1995)Dorschner, Begemann, Henning, J\"ager, \&
  Mutschke}]{DBH_95_glasses}
Dorschner, J., Begemann, B., Henning, T., J\"ager, C., \& Mutschke, H. 1995,
  \aap, 300, 503

\bibitem[{{Draine}(2003)}]{2003ApJ...598.1026D}
{Draine}, B.~T. 2003, \apj, 598, 1026

\bibitem[{{Draine} \& {Lee}(1984)}]{Draine84}
{Draine}, B.~T. \& {Lee}, H.~M. 1984, \apj, 285, 89

\bibitem[{{Egan} {et~al.}(2003){Egan}, {Price}, {Kraemer}, {Mizuno}, {Carey},
  {Wright}, {Engelke}, {Cohen}, \& {Gugliotti}}]{2003yCat.5114....0E}
{Egan}, M.~P., {Price}, S.~D., {Kraemer}, K.~E., {et~al.} 2003, VizieR Online
  Data Catalog, 5114

\bibitem[{{Elias}(1978)}]{Elias78}
{Elias}, J.~H. 1978, \apj, 224, 857

\bibitem[{{Evans} {et~al.}(2003){Evans}, {Allen}, {Blake}, {Boogert}, {Bourke},
  {Harvey}, {Kessler}, {Koerner}, {Lee}, {Mundy}, {Myers}, {Padgett},
  {Pontoppidan}, {Sargent}, {Stapelfeldt}, {van Dishoeck}, {Young}, \&
  {Young}}]{2003PASP..115..965E}
{Evans}, II, N.~J., {Allen}, L.~E., {Blake}, G.~A., {et~al.} 2003, \pasp, 115,
  965

\bibitem[{{Gillett} {et~al.}(1975){Gillett}, {Jones}, {Merrill}, \&
  {Stein}}]{1975A&A....45...77G}
{Gillett}, F.~C., {Jones}, T.~W., {Merrill}, K.~M., \& {Stein}, W.~A. 1975,
  \aap, 45, 77

\bibitem[{{Gustafsson} {et~al.}(1975){Gustafsson}, {Bell}, {Eriksson}, \&
  {Nordlund}}]{Gustafsson75}
{Gustafsson}, B., {Bell}, R.~A., {Eriksson}, K., \& {Nordlund}, A. 1975, \aap,
  42, 407

\bibitem[{{Gustafsson} {et~al.}(2008){Gustafsson}, {Edvardsson}, {Eriksson},
  {J{\o}rgensen}, {Nordlund}, \& {Plez}}]{2008A&A...486..951G}
{Gustafsson}, B., {Edvardsson}, B., {Eriksson}, K., {et~al.} 2008, \aap, 486,
  951

\bibitem[{Henning \& Stognienko(1996)}]{HS_96_opacities}
Henning, T. \& Stognienko, R. 1996, \aap, 311, 291

\bibitem[{{Houck} {et~al.}(2004){Houck}, {Roellig}, {van Cleve}, {Forrest},
  {Herter}, {Lawrence}, {Matthews}, {Reitsema}, {Soifer}, {Watson}, {Weedman},
  {Huisjen}, {Troeltzsch}, {Barry}, {Bernard-Salas}, {Blacken}, {Brandl},
  {Charmandaris}, {Devost}, {Gull}, {Hall}, {Henderson}, {Higdon}, {Pirger},
  {Schoenwald}, {Sloan}, {Uchida}, {Appleton}, {Armus}, {Burgdorf},
  {Fajardo-Acosta}, {Grillmair}, {Ingalls}, {Morris}, \&
  {Teplitz}}]{2004ApJS..154...18H}
{Houck}, J.~R., {Roellig}, T.~L., {van Cleve}, J., {et~al.} 2004, \apjs, 154,
  18

\bibitem[{Jones \& Merrill(1976)}]{JM_76_dust}
Jones, T.~W. \& Merrill, K.~M. 1976, \apj, 209, 509

\bibitem[{Kemper {et~al.}(2002)Kemper, {de Koter}, Waters, Bouwman, \&
  Tielens}]{KDW_02_composition}
Kemper, F., {de Koter}, A., Waters, L. B. F.~M., Bouwman, J., \& Tielens, A. G.
  G.~M. 2002, \aap, 384, 585

\bibitem[{{Kemper} {et~al.}(2004){Kemper}, {Vriend}, \& {Tielens}}]{Kemper04}
{Kemper}, F., {Vriend}, W.~J., \& {Tielens}, A.~G.~G.~M. 2004, \apj, 609, 826

\bibitem[{{Koornneef}(1983)}]{Koornneef83}
{Koornneef}, J. 1983, \aap, 128, 84

\bibitem[{{Mathis} {et~al.}(1977){Mathis}, {Rumpl}, \& {Nordsieck}}]{Mathis77}
{Mathis}, J.~S., {Rumpl}, W., \& {Nordsieck}, K.~H. 1977, \apj, 217, 425

\bibitem[{{McCarthy} {et~al.}(1980){McCarthy}, {Forrest}, {Briotta}, \&
  {Houck}}]{1980ApJ...242..965M}
{McCarthy}, J.~F., {Forrest}, W.~J., {Briotta}, Jr., D.~A., \& {Houck}, J.~R.
  1980, \apj, 242, 965

\bibitem[{{McClure}(2009)}]{2009ApJ...693L..81M}
{McClure}, M. 2009, \apjl, 693, L81

\bibitem[{{Min} {et~al.}(2005){Min}, {Hovenier}, \& {de
  Koter}}]{2005A&A...432..909M}
{Min}, M., {Hovenier}, J.~W., \& {de Koter}, A. 2005, \aap, 432, 909

\bibitem[{{Min} {et~al.}(2007){Min}, {Waters}, {de Koter}, {Hovenier},
  {Keller}, \& {Markwick-Kemper}}]{Min07}
{Min}, M., {Waters}, L.~B.~F.~M., {de Koter}, A., {et~al.} 2007, \aap, 462, 667

\bibitem[{Ossenkopf {et~al.}(1992)Ossenkopf, Henning, \&
  Mathis}]{OHM_92_cosmicsilicates}
Ossenkopf, V., Henning, T., \& Mathis, J.~S. 1992, \aap, 261, 567

\bibitem[{{Rawlings} {et~al.}(2000){Rawlings}, {Adamson}, \&
  {Whittet}}]{Rawlings00}
{Rawlings}, M.~G., {Adamson}, A.~J., \& {Whittet}, D.~C.~B. 2000, \apjs, 131,
  531

\bibitem[{{Rieke}(1974)}]{1974ApJ...193L..81R}
{Rieke}, G.~H. 1974, \apjl, 193, L81

\bibitem[{Roche(1988)}]{R_88_GC}
Roche, P.~F. 1988, in Dust in the universe, ed. M.~E. Bailey \& D.~A. Williams
  (Cambridge University Press), 415--433

\bibitem[{{Roche}(1989)}]{1989ESASP.290...79R}
{Roche}, P.~F. 1989, in ESA Special Publication, Vol. 290, Infrared
  Spectroscopy in Astronomy, ed. {E.~B{\"o}hm-Vitense}, 79--91

\bibitem[{{Roche} \& {Aitken}(1984)}]{Roche84}
{Roche}, P.~F. \& {Aitken}, D.~K. 1984, \mnras, 208, 481

\bibitem[{{Roche} \& {Aitken}(1985)}]{Roche85}
{Roche}, P.~F. \& {Aitken}, D.~K. 1985, \mnras, 215, 425

\bibitem[{{Rom{\'a}n-Z{\'u}{\~n}iga} {et~al.}(2007){Rom{\'a}n-Z{\'u}{\~n}iga},
  {Lada}, {Muench}, \& {Alves}}]{2007ApJ...664..357R}
{Rom{\'a}n-Z{\'u}{\~n}iga}, C.~G., {Lada}, C.~J., {Muench}, A., \& {Alves},
  J.~F. 2007, \apj, 664, 357

\bibitem[{{Scappini} {et~al.}(2003){Scappini}, {Cecchi-Pestellini}, {Smith},
  {Klemperer}, \& {Dalgarno}}]{2003MNRAS.341..657S}
{Scappini}, F., {Cecchi-Pestellini}, C., {Smith}, H., {Klemperer}, W., \&
  {Dalgarno}, A. 2003, \mnras, 341, 657

\bibitem[{{Schultz} \& {Wiemer}(1975)}]{1975A&A....43..133S}
{Schultz}, G.~V. \& {Wiemer}, W. 1975, \aap, 43, 133

\bibitem[{{Skrutskie} {et~al.}(2006){Skrutskie}, {Cutri}, {Stiening},
  {Weinberg}, {Schneider}, {Carpenter}, {Beichman}, {Capps}, {Chester},
  {Elias}, {Huchra}, {Liebert}, {Lonsdale}, {Monet}, {Price}, {Seitzer},
  {Jarrett}, {Kirkpatrick}, {Gizis}, {Howard}, {Evans}, {Fowler}, {Fullmer},
  {Hurt}, {Light}, {Kopan}, {Marsh}, {McCallon}, {Tam}, {Van Dyk}, \&
  {Wheelock}}]{2MASS06}
{Skrutskie}, M.~F., {Cutri}, R.~M., {Stiening}, R., {et~al.} 2006, \aj, 131,
  1163

\bibitem[{{Sneden} {et~al.}(1978){Sneden}, {Gehrz}, {Hackwell}, {York}, \&
  {Snow}}]{1978ApJ...223..168S}
{Sneden}, C., {Gehrz}, R.~D., {Hackwell}, J.~A., {York}, D.~G., \& {Snow},
  T.~P. 1978, \apj, 223, 168

\bibitem[{{Speck} {et~al.}(2008){Speck}, {Whittington}, \&
  {Tartar}}]{2008ApJ...687L..91S}
{Speck}, A.~K., {Whittington}, A.~G., \& {Tartar}, J.~B. 2008, \apjl, 687, L91

\bibitem[{{Stein} \& {Gillett}(1971)}]{1971Natur.233...72S}
{Stein}, W.~A. \& {Gillett}, F.~C. 1971, \nat, 233, 72

\bibitem[{{Strom} {et~al.}(1976){Strom}, {Vrba}, \&
  {Strom}}]{1976AJ.....81..314S}
{Strom}, S.~E., {Vrba}, F.~J., \& {Strom}, K.~M. 1976, \aj, 81, 314

\bibitem[{{Tsuji} {et~al.}(1994){Tsuji}, {Ohnaka}, {Hinkle}, \&
  {Ridgway}}]{1994A&A...289..469T}
{Tsuji}, T., {Ohnaka}, K., {Hinkle}, K.~H., \& {Ridgway}, S.~T. 1994, \aap,
  289, 469

\bibitem[{{van Dishoeck}(2004)}]{Dishoeck04}
{van Dishoeck}, E.~F. 2004, \araa, 42, 119

\bibitem[{{Whittet}(2003)}]{Whittet03}
{Whittet}, D.~C.~B. 2003, {Dust in the Galactic Environment}, (Bristol:
  Institute of Physics (IOP) Publishing)

\bibitem[{{Whittet} {et~al.}(1988){Whittet}, {Bode}, {Longmore}, {Adamson},
  {McFadzean}, {Aitken}, \& {Roche}}]{Whittet88_2}
{Whittet}, D.~C.~B., {Bode}, M.~F., {Longmore}, A.~J., {et~al.} 1988, \mnras,
  233, 321

\end{thebibliography}

\end{document}